%% file: aklt.tex
\documentclass[preprint,1p,12pt,number,sort&compress]{elsarticle}

\usepackage[T1]{fontenc}
\usepackage[utf8]{inputenc}
\usepackage[english]{babel}
\addto\captionsenglish{}
\usepackage{babel}
\usepackage{amsmath}
\usepackage{amssymb}
\usepackage{mathrsfs}
\usepackage{dcolumn}
\usepackage{bm}
\usepackage{braket}
\usepackage{graphicx}
\usepackage{comment}
\usepackage{hyperref}
\usepackage{color}
\usepackage{epsfig,natbib}
\usepackage{braket}
\usepackage{young}
\usepackage{youngtab}
\usepackage{ytableau}
\usepackage{relsize}
\usepackage{dsfont}
\usepackage{adjustbox}
\usepackage{relsize}
\usepackage{calc,ragged2e}
\usepackage{hyperref}
\hypersetup{
	colorlinks=true,
	linkcolor=blue
}
\usepackage{xcolor}
\usepackage{tikz}
\usetikzlibrary{decorations}
\usetikzlibrary{decorations.pathmorphing}
\usetikzlibrary{decorations.pathreplacing}
\usetikzlibrary{decorations.markings}
\usetikzlibrary{shapes.misc}
\usetikzlibrary{calc}
\def\mysquare[#1](#2)(#3){\draw[#1] (#2)  rectangle ($(#2)+({#3},{#3})$);}
\def\myrectangle[#1](#2)(#3){\draw[#1] (#2)  rectangle ($(#2)+({#3},{#3})$);\draw[#1] (#2)  rectangle ($(#2)+({#3},{2*#3})$);}
\newcolumntype{M}[1]{>{\centering\arraybackslash}m{#1}}
\newcolumntype{N}{@{}m{0pt}@{}}
\def \Tr {\text{Tr}}
\def \AKLT {\text{AKLT}}
\def \MPS {\text{MPS}}
\def \vir {\mathcal{V}}
\def \phys {\mathcal{P}}
\newcommand{\SU}[1]{\mathrm{SU}({#1})}
\newcommand{\yset}[1]{\mathrm{\bold{#1}}}
\newcommand{\be}{\begin{equation}}
\newcommand{\ee}{\end{equation}}
\newcommand{\bea}{\begin{eqnarray}}
\newcommand{\eea}{\end{eqnarray}}
\newcommand{\shsc}[3]{\mathrel{\raisebox{#1}{\protect\scalebox{#2}{#3}}}}

\journal{Nuclear Physics B}

\begin{document}

\begin{frontmatter}



\title{Novel families of $\SU{N}$ AKLT states with arbitrary self-conjugate edge states}


\author[epfl]{Samuel Gozel\corref{cor1}}
\ead{samuel.gozel@epfl.ch}
\author[irsamc,epfl]{Didier Poilblanc}
\author[ubc]{Ian Affleck}
\author[epfl]{Fr\'ed\'eric Mila}

\address[epfl]{Institute of Physics, \'Ecole Polytechnique F\'ed\'erale de Lausanne (EPFL), CH-1015 Lausanne, Switzerland}
\address[irsamc]{Laboratoire de Physique Th\'eorique, IRSAMC, Universit\'e de Toulouse, CNRS, UPS, France}
\address[ubc]{Department of Physics and Astronomy and Stewart Blusson Quantum Matter Institute, University of British Columbia, 
Vancouver, B.C., Canada, V6T1Z1}

\cortext[cor1]{Corresponding author}

\begin{abstract}
Using the Matrix Product State framework, we generalize the Affleck-Kennedy-Lieb-Tasaki (AKLT) construction to one-dimensional spin liquids with global color $\SU{N}$ symmetry, finite correlation lengths, and edge states that can belong to any self-conjugate irreducible representation (irrep) of $\SU{N}$. In particular, $\SU{2}$ spin-$1$ AKLT states with edge states of arbitrary spin $s=1/2,1,3/2,\cdots$ are constructed, and a general formula for their correlation length is given. Furthermore, we show how to construct local parent Hamiltonians for which these AKLT states are unique ground states. This enables us to study the stability of the edge states by interpolating between exact AKLT Hamiltonians. As an example, in the case of spin-$1$ physical degrees of freedom, it is shown that a quantum phase transition of central charge $c = 1$ separates the Symmetry Protected Topological (SPT) phase with spin-$1/2$ edge states from a topologically trivial phase with spin-$1$ edge states. We also address some specificities of the generalization to $\SU{N}$ with $N>2$, in particular regarding the construction of parent Hamiltonians. For the AKLT state of the $\SU{3}$ model with the $3$-box symmetric representation, we prove that the edge states are in the $8$-dimensional adjoint irrep, and for the $\SU{3}$ model with adjoint irrep at each site, we are able to construct two different reflection-symmetric AKLT Hamiltonians, each with a unique ground state which is either even or odd under reflection symmetry and with edge states in the adjoint irrep. Finally, examples of two-column and adjoint physical irreps for $\SU{N}$ with $N$ even and with edge states living in the antisymmetric irrep with $N/2$ boxes are given, with a conjecture about the general formula for their correlation lengths.
\end{abstract}

\begin{keyword}
AKLT \sep $\SU{N}$ symmetry \sep spin chain \sep matrix product state


\end{keyword}

\end{frontmatter}

\section{Introduction}

The Affleck-Kennedy-Lieb-Tasaki (AKLT) model of a spin-$1$ chain~\cite{affleck_rigorous_1987,affleck_valence_1988}, with a biquadratic interaction equal to a third of the bilinear one, has played an important role in proving Haldane's conjecture that the Heisenberg spin-$1$ chain is gapped~\cite{haldane_nonlinear_1983,haldane_continuum_1983}. Indeed, it is the first model in the Haldane phase of the bilinear-biquadratic spin-$1$ chain for which it could be proven analytically that the spectrum is gapped~\cite{affleck_rigorous_1987,affleck_valence_1988}. This result was all the more important given that the other points for which exact results were known, with a biquadratic interaction equal or opposite to the bilinear interaction, were known to be gapless~\cite{babujian_exact_1982,takhtajan_picture_1982,uimin_one_1970,lai_lattice_1974,sutherland_model_1975}. 

Another equally important aspect of the AKLT construction is its exact ground state, the first example of an exact wave function realizing a gapped $\SU{2}$-symmetric spin liquid hosting protected edge states at the end of open chains. In the AKLT construction the physical spins are written in terms of two virtual spin-$1/2$ degrees of freedom attached to each lattice site which, simultaneously, form maximally entangled bond singlets between neighboring sites. This can be conveniently reformulated in terms of a Matrix Product State (MPS), and in that respect the AKLT construction has played an important role in popularizing the MPS, which is nowadays the standard formulation of the Density Matrix Renormalization Group (DMRG)~\cite{white_density_1992,white_density_1993,ostlund_thermodynamic_1995,schollwock_density_2011}. 

In addition to the well-known spin-$1$ AKLT chain, AKLT introduced several other valence-bond solid (VBS) states in one and two dimensions (the construction beeing the same in any dimension), always using virtual spin-$1/2$ degrees of freedom, but either taking lattices with higher coordination number $z$ such that the physical spin $S$ satisfies $S = z/2$ (for instance $S=3/2$ on the honeycomb lattice or $S=2$ on the square lattice) or also by forming an integer number $n$ of singlets on each bond, such that the physical spin is now $S = n z/2$~\cite{affleck_rigorous_1987,affleck_valence_1988}. The AKLT construction has then been generalized to integer spin-$S$ chains with $S/2$ edge states~\cite{arovas_extended_1988}, and to VBS with $\SU{2N}$ symmetry but breaking translation and charge conjugation symmetries~\cite{affleck_su2n_1991}. More recently, a number of generalizations have been proposed for $\SU{N}$ chains as well~\cite{greiter_exact_2007,greiter_valence_2007,katsura_entanglement_2008,morimoto_z_2014,nonne_symmetry-protected_2013,roy_chiral_2018}. In all these constructions, the physical state is built as a composite object of irreps of smaller dimension, and the edge states are ``fractionalized''\cite{chen_classification_2011,schuch_classifying_2011}.

In the present paper, we discuss another generalization of the AKLT construction in which the physical state is built as a composite object of irreps of {\it arbitrary} dimension. The only condition on these irreps is that they should be self-conjugate so that two of them can be used to build a singlet. This construction is most simply done in the MPS language, which we use throughout, and parent Hamiltonians are constructed in a systematic way. This allows us for instance to discuss models of spin-$1$ chains with edge states of arbitrary spin $s = 1/2,1,3/2,\cdots$, and to study the quantum phase transition between topological and trivial phases with half-integer respectively integer spin edge states. This construction turns out to be quite useful for $\SU{N}$ states as well. It allows one to prove for instance that the AKLT state for the $3$-box symmetric representation has edge states in the $8$-dimensional adjoint irrep, a result plausible but not proven in the context of the construction in terms of three local fundamental representations. 

The paper is organized as follows. In Section~\ref{sec::general_AKLT}, we describe the general construction of AKLT states with arbitrary self-conjugate edge states, and we discuss the case of $\SU{2}$ spin-$S$ states with $S$ integer. Section~\ref{sec::parentHamiltonians} is devoted to the construction of parent Hamiltonians, with the spin-$1$ chain as an example. In Section~\ref{sec::interpolspin1}, we study a Hamiltonian that interpolates between the original AKLT model and one of the parent Hamiltonians of the spin-$1$ AKLT state with spin-$1$ edge states, and we show that it undergoes a quantum phase transition with central charge $c = 1$. In Section~\ref{sec::SUN}, we discuss several aspects of the construction for $\SU{N}$ models, with emphasis on $\SU{3}$ examples. Finally the results are summarized in Section~\ref{sec::conclusion}.

\section{General construction of AKLT states}
\label{sec::general_AKLT}

We consider a one-dimensional lattice of $L$ ``spins'' whose local degrees of freedom are described by an irreducible representation (irrep) $\phys$ of a continous group $G$. In what follows we will mainly be concerned with the case of $G = \SU{N}$ where the symmetry properties of the irrep $\mathcal{P}$ are uniquely determined by its Young tableau $\alpha^{\phys}$, an array of boxes with $1 \leqslant k< N$ rows such that the lengths $\alpha^{\phys}_j$ of the $j$\textsuperscript{th} row satisfy $\alpha^{\phys}_1 \geqslant \alpha^{\phys}_2 \geqslant ... \geqslant \alpha^{\phys}_k > 0$. We denote this Young tableau by $\alpha^{\phys} = [\alpha^{\phys}_1, \, \alpha^{\phys}_2, ... ,\, \alpha^{\phys}_k]$. When it does not lead to extra confusion $\phys$ and $\alpha^{\phys}$ will be used interchangeably to denote the irrep $\phys$ of $G$ (but special caution must be brought to the case where non-trivial outer multiplicities occur\footnote{The {\it outer multiplicity} of a given irrep in the tensor product of two irreps is defined as the number of times it appears when writing this tensor product as a direct sum of irreps. The {\it inner multiplicity} of a pattern-weight (p-weight) corresponds to the number of states of a given irrep having the same pattern-weight, namely the same eigenvalues in the Cartan subalgebra~\cite{alex_numerical_2011}.}, see below).

\begin{figure}
\begin{minipage}[c][26mm]{1.0\textwidth}
\input{fig_aklt_state}
\end{minipage}
\begin{minipage}[c][30mm]{0.45\textwidth}
\input{fig_CGC}
\end{minipage}
\begin{minipage}[c][30mm]{0.55\textwidth}
\input{fig_aklt_state_mps}
\end{minipage}
\hfill
\begin{minipage}[c][40mm]{0.5\textwidth}
\input{fig_aklt_ortho}
\end{minipage}
\begin{minipage}[c][40mm]{0.5\textwidth}
\input{fig_mps_H}
\end{minipage}
\caption{(a)~AKLT state $\ket{\phys, \vir}$. The ellipses denote physical sites where two virtual irreps $\vir$ represented by the black dots are mapped onto the physical irrep $\phys$. Singlets on each bond are represented by a thick line joining adjacent virtual irreps from adjacent sites. On an open chain the edge states are determined by the unpaired virtual spins $\vir$ at the left and right ends of the chain. (b)~Clebsch-Gordan coefficient $\braket{\phys, \mathfrak{h}_{\sigma}| \vir_1, \mathfrak{f}_a; \vir_2, \mathfrak{g}_b}$. (c)~MPS form for PBC. The black circle is the singlet matrix and a wiggly line denotes a physical index. (d)~$\ket{\phys, \vir, \nu; (\alpha, \mathfrak{h}^{\alpha}_{\sigma}, j)}$. The thick leg pointing downward means that all relevant irreps $\alpha$ with states $\mathfrak{h}^{\alpha}_{\sigma}$ are kept. (e)~Parent Hamiltonian.}
\label{fig::aklt}
\end{figure}
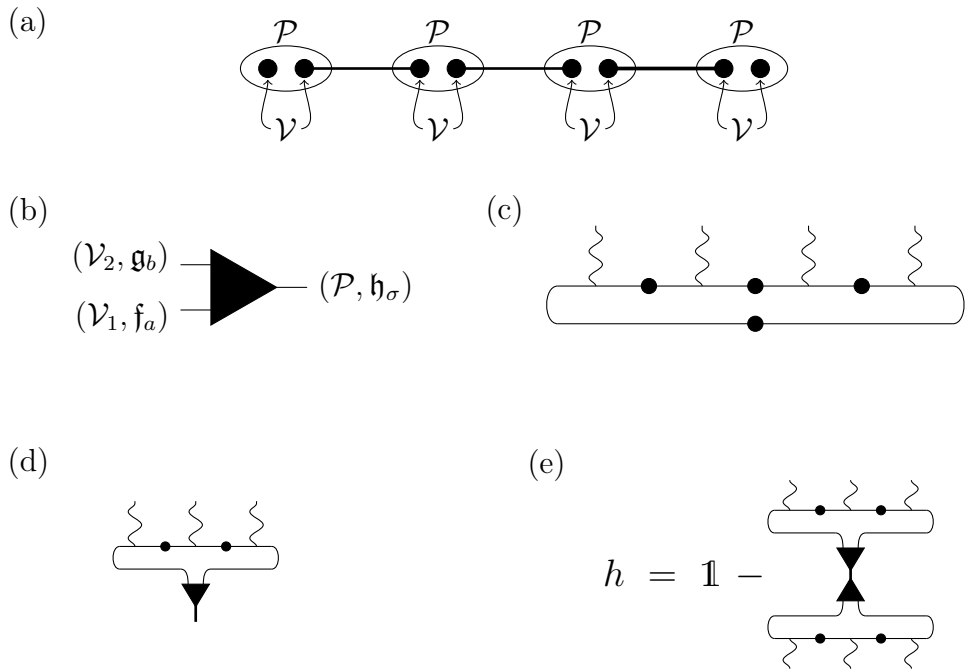

Any state of the one-dimensional lattice with $L$ sites is a state living in the tensor product space $\phys^{\otimes L}$. 
An AKLT state, or VBS, with physical irrep $\phys$ on each site is a state of $\phys^{\otimes L}$ formed with singlets on each bond between nearest neighbor sites. This state can be built by decomposing the physical spin on each site into two ``virtual'' spins described by the irreps $\vir_L$ and $\vir_R$ of $G$ such that $\phys \in \vir_L \otimes \vir_R$. The singlets on each bond are then obtained by mapping two virtual spins $\vir_R$, $\vir_L$ from neighboring sites onto the singlet irrep and by mapping the two virtual spins on each physical site onto the physical irrep. The virtual spins $\vir_L$, $\vir_R$ determine the nature of the edge states at the left and right end of a finite chain with open boundary conditions (OBC). Since the singlet must belong to $\vir_L \otimes \vir_R$, the two virtual spins must be conjugate: $\vir_L^* = \vir_R$. When $\vir_L \neq \vir_R$ one can construct two AKLT states corresponding to the two different ways of forming a singlet on two neighboring sites (this is obtained by exchanging $\vir_L$ and $\vir_R$). These two states break inversion symmetry. In order to avoid this unwanted degeneracy we focus here on self-conjugate virtual spins $\vir \equiv \vir_L = \vir_R$. This defines an AKLT state $\ket{\AKLT} \equiv \ket{\phys, \vir}$, Fig.~\ref{fig::aklt}(a). 
When considering $G = \SU{N}$ with $N>2$ the construction above needs to be further specified because non-trivial outer multiplicities can appear in the tensor product of two virtual spins  
\be
\mathcal{U} = \vir\otimes\vir.
\ee 
An example is provided in Fig.~\ref{fig::su3::p[300]::v[210]}(a) in the case of $G=\SU{3}$ and $\vir=\ \shsc{2pt}{0.4}{\ydiagram{2,1}} $.
The space $\mathcal{U}$ can be described by its content in terms of Young tableaux $\yset{U}$,
\be
\mathcal{U} = \underset{\alpha\in\yset{U}}{\mathlarger{\mathlarger{\oplus}}} \, U^{\alpha}.
\ee
Fig.~\ref{fig::su3::p[300]::v[210]}(b) shows the set $\yset{U}$ in the above $\SU{3}$ example. The singlet irrep $\alpha \equiv \bullet$ appears necessarily in $\yset{U}$ with outer multiplicity $1$, because $\vir$ is self-conjugate. This is unlike the physical irrep for which the shape $\alpha^{\phys} \in \yset{U}$ can appear with outer multiplicity 
$\mu^{\alpha^{\phys}}_{\mathcal{U}} \equiv \mu^{\phys} > 1$ in $\mathcal{U}$,
as shown e.g. in Fig.~\ref{fig::su3::p[300]::v[210]} for $\phys= \ \shsc{2pt}{0.4}{\ydiagram{2,1}}$. 
Each of the $\mu_{\phys}$ copies of $\phys$ in $\mathcal{U}$ leads to an AKLT state, which we denote by
\be
\ket{\phys, \vir, \nu}, \quad \nu=1,2,...,\mu_{\phys}.
\ee
Note that $\mathcal{U} = \vir\otimes\vir$ also defines the Hilbert space associated with the edge modes of an open chain.
As a consequence we observe that once a virtual irrep is chosen, namely once the edge states of an open chain are determined, AKLT states 
possessing these edge states can be built for different physical spins.

\begin{figure}
\begin{minipage}[c][16mm]{1.0\textwidth}
\input{fig_su3_adjkronadj2}
\end{minipage}
\begin{minipage}[c][16mm]{1.0\textwidth}
\input{fig_su3_content_irreps2}
\end{minipage}
\caption{(a) Decomposition of ${\cal U}=\vir\otimes\vir$ in the case of the $\SU{3}$ AKLT states $\ket{\phys, \vir = \ \shsc{2pt}{0.4}{\ydiagram{2,1}} \, }$. 
(b) Irrep content $\yset{U}$ of the product subspace $\cal{U}$. For $\phys=  \ \shsc{0pt}{0.4}{\ydiagram{3}}$ ($\phys= \ \shsc{2pt}{0.4}{\ydiagram{2,1}}$) a unique (two) AKT state(s) can be constructed.  }
\label{fig::su3::p[300]::v[210]}
\end{figure}

\subsection{AKLT states as MPS}

\label{sec::aklt_mps}

The construction of the AKLT states described above can be made explicit by use of the Clebsch-Gordan coefficients (CGC) for the irreps of the group $G$. For two arbitrary irreps $\vir_1, \vir_2$ of $G$ and an irrep $\phys \in \vir_1 \otimes \vir_2$ we define the CGC associated with $\vir_1 \otimes \vir_2 \rightarrow \phys$ as
\be
\braket{\phys, \mathfrak{h}_{\sigma}| \vir_1, \mathfrak{f}_a; \vir_2, \mathfrak{g}_b}, \quad
\begin{array}{rcl}
a & = & 1,..., \dim(\vir_1) \\
b & = & 1,..., \dim(\vir_2) \\
\sigma & = & 1,..., \dim(\phys), \\
\end{array}
\ee
where $\mathfrak{f}_a$, $\mathfrak{g}_b$ and $\mathfrak{h}_{\sigma}$ identify uniquely the states. For instance when $G = \SU{N}$ one choses $\mathfrak{f}_a$, $\mathfrak{g}_b$ and $\mathfrak{h}_{\sigma}$ to be the eigenvalues of the associated states in the Cartan subalgebra. Figure~\ref{fig::aklt}(b) is a useful pictorial representation of the CGC as a $2$-to-$1$ fusion tensor.
For a self-conjugate virtual irrep $\vir$ we define the CGC associated with the singlet
\be
S_{a,b} = \braket{\bullet| \vir, \mathfrak{f}_a; \vir, \mathfrak{f}_b}
\ee
and the ones associated with the physical spins,
\be
M^{\sigma,\nu}_{a,b} = \braket{\phys, \mathfrak{h}_{\sigma}, \nu| \vir, \mathfrak{f}_a; \vir, \mathfrak{f}_b}, \quad \nu=1,...,\mu_{\phys}.
\ee
The $\mu_{\phys}$ (unnormalized) AKLT states on a periodic chain of length $L$ are then given by the Matrix Product States (MPS)
\be
\ket{\phys, \vir, \nu} = \sum_{\bm{\sigma}} \Tr\left( M^{\sigma_1,\nu} S M^{\sigma_2,\nu} S ... M^{\sigma_L,\nu} S\right) \ket{\bm{\sigma}}.
\ee
The pictorial representation of this MPS is given in Fig.~\ref{fig::aklt}(c).
In what follows we call $d = \dim(\phys)$ and $D = \dim(\vir)$, the dimensions of the physical and virtual spaces, respectively. 
Defining (we temporarily forget about the outer  multiplicity index $\nu$)
\be
A^{\sigma} := M^{\sigma} S
\ee
the correlation length $\xi$ of the state $\ket{\phys, \vir}$ is directly accessible from the two eigenvalues of largest real part $\lambda_1, \lambda_2$ of the transfer matrix\footnote{Since the CGC can be chosen to be real, the transfer matrix is real by definition.}
\be
\mathcal{T} := \sum_{\sigma} A^{\sigma} \otimes A^{\sigma}
\ee
through
\be
\xi^{-1} = \ln\left(\left|\frac{\lambda_1}{\text{Re} \, \lambda_2}\right|\right).
\ee
For a non-trivial outer multiplicity $\mu_{\phys}$ the correlation length $\xi$ can be computed for each state $\ket{\phys, \vir, \nu}$.

\subsection{Examples: $\SU{2}$ integer spin-$S$ AKLT states}

The original AKLT state is a spin-$1$ chain made of virtual spin-$1/2$ degrees of freedom~\cite{affleck_rigorous_1987,affleck_valence_1988}. From the  discussion above one sees that there is no such AKLT state with spin-$1/2$ edge states for higher values of the physical spin. More generally one can construct an $\SU{2}$ AKLT state with physical integer spin-$S$ with any virtual spin-$s$ satisfying $s\geqslant S/2$. This more general construction was introduced initially in Ref.~\cite{tu_topologically_2009}. We focus here on the construction of a spin-$1$ AKLT state with spin-$1$ edge states to illustrate the procedure.  Taking the eigenvalues in the Cartan subalgebra $S^z$ with the labelling $\mathfrak{f}_1 = +1, \mathfrak{f}_2 = 0, \mathfrak{f}_3 = -1$, the necessary CGC are given by~\cite{alex_numerical_2011},
\be
S = \frac{1}{\sqrt{3}}\begin{pmatrix}
0 & 0 & 1 \\
0 & -1 & 0 \\
1 & 0 & 0\\
\end{pmatrix},
\ee

\be
\begin{aligned}
M^{1} & = \frac{1}{\sqrt{2}}\begin{pmatrix}
0 & -1 & 0 \\
1 & 0 & 0 \\
0 & 0 & 0 \\
\end{pmatrix}, \ 
M^{2} = \frac{1}{\sqrt{2}}\begin{pmatrix}
0 & 0 & -1 \\
0 & 0 & 0 \\
1 & 0 & 0 \\
\end{pmatrix}, \
M^{3} = \frac{1}{\sqrt{2}}\begin{pmatrix}
0 & 0 & 0 \\
0 & 0 & -1 \\
0 & 1 & 0 \\
\end{pmatrix}.
\end{aligned}
\ee
One obtains the correlation length $\xi = 1/\ln 2$ for this AKLT state, to be compared to $\xi = 1/\ln 3$ for the original AKLT state with spin-$1/2$ edge states~\cite{tu_topologically_2009}. For a physical spin $S = 1$ and a virtual spin-$s$ we conjecture the following expression for the correlation length,
\be
\xi = \frac{1}{\ln\left| 1 + \frac{1}{s(s+1)-1}\right|}
\label{eq::xi1}
\ee
which increases as $\xi \simeq s (s+1)$ for large $s$. More generally for a physical integer spin-$S$ with a virtual spin $s=S/2$ or $s>S$ one obtains
\be
\xi = - \frac{1}{\ln\left| 1 - \frac{C_2(S)}{2 C_2(s)}\right|}
\label{eq::xi2}
\ee
where $C_2(S) = S(S+1)$ is the eigenvalue of the quadratic Casimir operator. For $s$ large the correlation length increases as $\xi \simeq 2 C_2(s)/C_2(S)=2s(s+1)/S(S+1)$. The increase of the correlation length as the virtual spin grows is probably related to the increase of the number of sites required to write a valid parent Hamiltonian (see below). Equations~\eqref{eq::xi1} and~\eqref{eq::xi2} have been empirically established on the basis of 
numerical calculations of the correlation length for $S = 1,2,3,4$ and $s = S,S+1/2, S+1$, but can be proved rigorously~\cite{tu_private_2019}.

\section{Parent Hamiltonians}
\label{sec::parentHamiltonians}

Since the work of AKLT parent Hamiltonians for VBS states were constructed using projectors defined with the use of the quadratic Casimir operator~\cite{affleck_rigorous_1987,affleck_valence_1988,greiter_exact_2007,greiter_valence_2007,tu_topologically_2009,nonne_symmetry-protected_2013,morimoto_z_2014}. This method, in spite of its extreme simplicity, has the disadvantage that it does not always lead to a parent Hamiltonian with a unique ground state, as we will show below. This failure was circumvented by a more sophisticated procedure introduced in Ref.~\cite{schuch_classifying_2011} and~\cite{roy_chiral_2018}, where the MPS form of the ground state is explicitly exploited, only requiring that the MPS is injective (see below). Here we simply revisit the construction in a more practical way and show how one can extract an explicit expression for the Hamiltonians in terms of spin operators, making the $\SU{N}$ symmetry manifest.

\subsection{AKLT construction}

Parent Hamiltonians are defined as Hamiltonians for which a specific state is the unique ground state (GS). In our case we are looking for local gapped parent Hamiltonians which have a unique ground state for PBC and a $\mathcal{E}$-dimensional ground state manifold for OBC, where $\mathcal{E}=D^2$. Originally the parent Hamiltonian of the standard AKLT state was constructed as a sum of projectors onto the subspace of total spin-$2$ on two neighboring sites,
\be
\mathcal{H}_{\AKLT} = 2 \sum_{i} \mathbb{P}^{S=2}_{i,i+1},
\ee
and, after subsequent rewriting of the projectors in terms of spin operators~\citep{affleck_rigorous_1987},
\be
\mathcal{H}_{\AKLT} = \sum_i \left[ \bold{S}_i \cdot \bold{S}_{i+1} + \frac{1}{3} (\bold{S}_i \cdot \bold{S}_{i+1})^2 + \frac{2}{3} \right].
\label{equ::H_AKLT_87}
\ee
This construction ensures that all states having spin $0$ or $1$ on two neighboring sites have vanishing energy, while spin-$2$ states on two neighboring sites are given a strictly positive energy. It was shown later that this Hamiltonian remains gapped in the thermodynamic limit~\citep{affleck_valence_1988}. The obvious problem with this method is that it does not always lead to a Hamiltonian with unique GS. For instance let us consider the AKLT state $\ket{S=1, s=1}$. Since $\vir=\phys$ a two-site approach would lead to the trivial solution $\mathcal{H} = 0$, because $\mathcal{P}^{\otimes 2}  \setminus \mathcal{U} = \emptyset$, namely there isn't any state which belongs to $\mathcal{P}^{\otimes 2}$ but does not lie in $\mathcal{U}$. One must thus take $l>2$ sites to form the Hamiltonian. 
 
\subsection{Physical Hilbert space on $l$ sites}

\label{app:irrepsun}

 The full Hilbert space of $l\leqslant L$ spins $\phys$ 
can be decomposed into a direct sum of irreducible representations,
\be
\phys^{\otimes l} = \underset{\alpha\in\yset{A}}{\mathlarger{\mathlarger{\oplus}}} V^{\alpha}
\ee
where $\yset{A}$ is the set of Young tableaux $\alpha$ characterizing the transformation properties of the Hilbert space $V^{\alpha}$ under $\SU{N}$. The Hilbert space $V^{\alpha}$ can itself be decomposed into $\mu^{\alpha}$ Hilbert spaces $V^{\alpha}_i, \ i=1,...,\mu^{\alpha}$ if the Young tableau $\alpha$ appears with a non-trivial outer multiplicity $\mu^{\alpha}$ in the tensor product $(\alpha^{\phys})^{\otimes l}$.

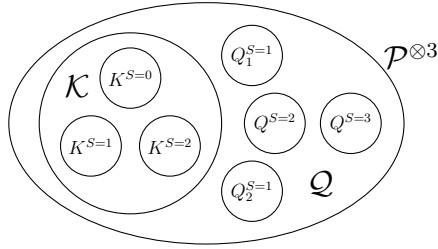
\begin{figure}
\begin{center}
\input{fig_spin1_sets.tex}
\end{center}
\caption{Decomposition of $\phys^{\otimes 3} \equiv (\text{spin-}1)^{\otimes 3}$ for the construction of the parent Hamiltonian of the AKLT state $\ket{S=1, s=1}$ on $l = 3$ sites. Here the space $\mathcal{U} := \vir\otimes\vir \subset \phys^{\otimes 3}$, thus $\mathcal{U} = \mathcal{K}$.}
\label{fig::su2spin1sets}
\end{figure}

The Hilbert space associated with the edge modes of a $l$-site open chain
$
\mathcal{U} = \vir\otimes\vir
$
is not necessarily entirely contained in $\phys^{\otimes l}$ which motivates the definition of
\be
\mathcal{K} = \phys^{\otimes l} \cap \mathcal{U} 
\ee
as well as its complement in $\phys^{\otimes l}$,
\be
\mathcal{Q} = \phys^{\otimes l} \setminus \mathcal{K}.
\ee
We should assume $\mathcal{K}$ and $\mathcal{Q}$ to be non-empty. If $\mathcal{Q} = \emptyset$ for a given $l$, it can be made non-trivial by increasing $l$. The spaces $\mathcal{U}$, $\mathcal{K}$ and $\mathcal{Q}$ can be described by their content in terms of Young tableaux $\yset{U}$, $\yset{K}$ and $\yset{Q}$, respectively,
\be
\mathcal{U} = \underset{\alpha\in\yset{U}}{\mathlarger{\mathlarger{\oplus}}} \, U^{\alpha}, \qquad \mathcal{K} = \underset{\alpha\in\yset{K}}{\mathlarger{\mathlarger{\oplus}}} \, K^{\alpha} \qquad \text{and} \qquad \mathcal{Q} = \underset{{\alpha\in\yset{Q}}}{\mathlarger{\mathlarger \oplus}} \, Q^{\alpha}.
\ee
Denoting by $\mu^{\alpha}_{\mathcal{K}}$ and $\mu^{\alpha}_{\mathcal{Q}}$ the outer multiplicities of the shape $\alpha$ one can further decompose $K^{\alpha}$ and $Q^{\alpha}$ as
\be
K^{\alpha} = \overset{\mu^{\alpha}_\mathcal{K}}{\underset{j=1}{\mathlarger{\mathlarger{\oplus}}}} \, K^{\alpha}_j, \ \alpha \in \yset{K}, \qquad 
Q^{\alpha} = \overset{\mu^{\alpha}_\mathcal{Q}}{\underset{j=1}{\mathlarger{\mathlarger{\oplus}}}} \, Q^{\alpha}_j, \ \alpha \in \yset{Q}.
\ee
Let us now come back to the AKLT state $\ket{S=1, s=1}$, for which a two-site approach is not valid, and let us consider $l=3$. Then,
\be
\bold{3}\otimes\bold{3}\otimes\bold{3} = \bold{1} \oplus 3 \times \bold{3} \oplus 2 \times \bold{5} \oplus \bold{7}
\label{equ::pS1vS1::tensor_prod_1}
\ee
where a spin-$S$ is denoted by its irrep dimension $\bold{2S+1}$. One would then infer the form of the Hamiltonian
\be
\mathcal{H} = \sum_i \mathbb{P}^{S=3}_{i,i+1,i+2}
\label{equ::pS1vS1::wrong_H}
\ee
where again the projector onto the spin-$3$ subspace can be expressed in terms of spin operators (see Eq.~\eqref{equ::su2::pS1::projector_spin3}). The Hamiltonian in Eq.~\eqref{equ::pS1vS1::wrong_H} however is not a valid parent Hamiltonian as defined above, because its GS is not unique for PBC. This can be easily understood by using the notations above. We have $\phys^{\otimes 3} = V^{S=0} \oplus V^{S=1} \oplus V^{S=2} \oplus V^{S=3}$ and the spin-$1$ subspace can be decomposed into $3$ parts while the spin-$2$ subspace can be decomposed into $2$ parts. Here $\mathcal{K} := \phys^{\otimes l} \cap \, \mathcal{U} = \mathcal{U}$ can be decomposed as $\mathcal{K} = K^{S=0} \oplus K^{S=1} \oplus K^{S=2}$ and $\mathcal{Q} := \phys^{\otimes l} \setminus \mathcal{K} = Q^{S=1}_1 \oplus Q^{S=1}_2 \oplus Q^{S=2} \oplus Q^{S=3}$, as illustrated in Fig.~\ref{fig::su2spin1sets}. The projector in Eq.~\eqref{equ::pS1vS1::wrong_H} annihilates all states in $V^{S=0} \oplus V^{S=1} \oplus V^{S=2}$, namely $\mathcal{K} \oplus Q^{S=1}_1 \oplus Q^{S=1}_2 \oplus Q^{S=2}$ while the correct kernel should be restricted to $\mathcal{K}$.

\subsection{Construction from MPS}

The MPS form of the AKLT state $\ket{\phys, \vir, \nu}$ on $l$ sites with open boundary conditions (OBC) can be viewed as a $D^2\times d^l$ linear map from the virtual unpaired states to the physical space,
\be
\begin{array}{lclcll}
\mathcal{M}^{\nu}: & \vir & \otimes & \vir & \rightarrow &  \phys^{\otimes l}\\
& \ket{\vir, \mathfrak{f}_a} & \otimes & \ket{\vir, \mathfrak{f}_b} & \rightarrow & \ket{\phys, \vir, \nu; \mathfrak{f}_a, \mathfrak{f}_b}.\\
\end{array}
\ee
Note that it is not assumed here that $\mathcal{K}=\mathcal{U}$, as in Fig.~\ref{fig::su2spin1sets}, but $\mathcal{K}\neq\mathcal{U}$ is also possible.
The state $\ket{\phys, \vir, \nu; \mathfrak{f}_a, \mathfrak{f}_b}$ is explicitly given by
\be
\ket{\phys, \vir, \nu; \mathfrak{f}_a, \mathfrak{f}_b} = \sum_{\bm{\sigma}} \left( M^{\sigma_1,\nu} S M^{\sigma_2,\nu} S ... M^{\sigma_l,\nu}\right)_{a,b} \ket{\bm{\sigma}}.
\ee
The MPS on $l$ sites is said to be injective if the map $\mathcal{M}^{\nu}$ is injective. Assuming now that the map $\mathcal{M}^{\nu}$ is injective (if it is not, the injectivity condition can be reached by increasing $l$~\cite{fannes_finitely_1992,nachtergaele_spectral_1996,perez-garcia_matrix_2006,sanz_quantum_2010,schuch_classifying_2011}), the $\mathcal{E}=D^2$ linearly independent AKLT states $\ket{\phys, \vir, \nu; \mathfrak{f}_a, \mathfrak{f}_b}$ are not all orthogonal to each other. Taking $\ket{\phys, \vir, \nu; \mathfrak{f}_a, \mathfrak{f}_b}$ to be unit-normalized one can form an orthonormal basis of the image of $\mathcal{M}^{\nu}$ by taking
\be
\begin{aligned}
& \ket{\phys, \vir, \nu; (\alpha, \mathfrak{h}^{\alpha}_{\sigma}, j)}
 = \sum_{a,b=1}^{D} \braket{\alpha, \mathfrak{h}^{\alpha}_{\sigma}, j| \vir, \mathfrak{f}_a; \vir, \mathfrak{f}_b} \ket{\phys, \vir, \nu; \mathfrak{f}_a, \mathfrak{f}_b},\\
& \hspace{1cm}\alpha\in\yset{U}, \ \sigma=1,...,\dim(U^{\alpha}_j), \  j = 1,2,...,\mu^{\alpha}_{\mathcal{U}}.\\
\end{aligned}
\label{equ::aklt_states_obc_normalized_mps}
\ee
This rotation is pictorially represented in Fig.~\ref{fig::aklt}(d) where we attach the $2$-to-$1$ fusion tensor of CGC to the MPS of the AKLT state. In Eq.~\eqref{equ::aklt_states_obc_normalized_mps} the irreps $\alpha$ with outer multiplicity $\mu^{\alpha}_{\mathcal{U}}$ and states labelled by their eigenvalues $\mathfrak{h}^{\alpha}_{\sigma}$ in the Cartan subalgebra are the ones which appear in the tensor product $\mathcal{U} = \vir\otimes\vir$. This change of basis ensures that all states are labelled by $\SU{N}$ indices with proper eigenvalues in the tensor product basis of the edge irreps. The states defined in Eq.~\eqref{equ::aklt_states_obc_normalized_mps} allow us to construct projectors onto the different components of $\mathcal{K}$ as
\be
\begin{aligned}
& \mathbb{P}^{\nu}_{K^{\alpha}_{j}} = \sum_{\sigma=1}^{\dim(K^{\alpha}_j)} \ket{\phys, \vir, \nu; (\alpha, \mathfrak{h}^{\alpha}_{\sigma}, j)}\bra{\phys, \vir, \nu; (\alpha, \mathfrak{h}^{\alpha}_{\sigma}, j)},\\
& \qquad \forall \alpha\in\yset{K}, \quad j=1,2,...,\mu^{\alpha}_{\mathcal{K}}.
\end{aligned}
\label{equ::mps_proj}
\ee
Defining the $l$-site local Hamiltonian given in Fig.~\ref{fig::aklt}(e)~\citep{roy_chiral_2018},
\be
h^{\nu} = \mathds{1} - \mathbb{P}^{\nu}_{\MPS}
\ee
where
\be
\mathbb{P}^{\nu}_{\MPS}\equiv\mathbb{P}^{\nu}_{\mathcal{K}} = \sum_{\alpha\in \yset{K}} \ \sum_{j=1}^{\mu^{\alpha}_{\mathcal{K}}} \mathbb{P}^{\nu}_{K^{\alpha}_{j}},
\ee
the parent Hamiltonian of the AKLT state $\ket{\phys, \vir, \nu}$ is finally given by
\be
\mathcal{H}^{\nu} = \sum_{i} \tau_i(h^{\nu})
\ee
where $\tau_i$ is the translation operator such that $\tau_i(h^{\nu})$ acts on sites $(i,i+1,...,i+l-1)$.

\subsection{Family of parent Hamiltonians}

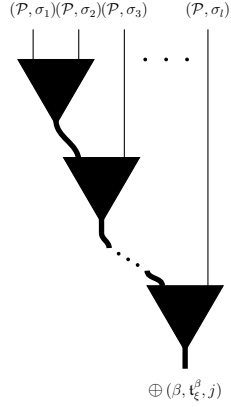
\begin{figure}
\input{fig_allstates_cgc}
\caption{Construction of the states of $\phys^{\otimes l}$ in definite symmetry sectors.}
\label{fig::states_cgc}
\end{figure}

Parent Hamiltonians of AKLT states are not unique. A simple example consists in applying the method presented above on the original AKLT state on $3$ sites. One would obtain a $3$-site local AKLT Hamiltonian, while the usual form is the $2$-site local Hamiltonian in Eq.~\eqref{equ::H_AKLT_87}. For a given length $l$ it is actually possible to determine the full family of parent Hamiltonians acting on at most $l$ sites. Let us consider again the physical space $\phys^{\otimes l}$ and let us assume that $\mathcal{K}=\mathcal{U}$, as in Fig.~\ref{fig::su2spin1sets}, so that $\mathbb{P}_{\mathcal{K}}+\mathbb{P}_{\mathcal{Q}}= \mathds{1}$. Using the CGC sequentially one can express all the states of each irrep $\beta \in \yset{A}$ in terms of the states of the tensor product basis. We denote these orthonormal states by
\be
\ket{\Psi_{(\beta, \mathfrak{t}^{\beta}_{\xi}, j)}}.
\ee
They correspond to contracting the tensor network given in Fig.~\ref{fig::states_cgc}. 
There is some arbitrariness in the CGC and in the labelling of the states. We exploit this arbitrariness to ensure that the AKLT states defined in Sec.~\ref{sec::aklt_mps} are a subset of the orthonormal basis $\left\{ \ket{\Psi_{(\beta, \mathfrak{t}^{\beta}_{\xi}, j)}} \right\}$. Projectors onto the different subsectors can be constructed as follows,
\be
\mathbb{P}^{\beta}_{j} = \sum_{\xi} \ket{\Psi_{(\beta, \mathfrak{t}^{\beta}_{\xi}, j)}}\bra{\Psi_{(\beta, \mathfrak{t}^{\beta}_{\xi}, j)}}.
\label{equ::proj_1}
\ee
Similarly one can define ``intertwiners'' which map states of one irrep to the other,
\be
\mathbb{X}^{\beta}_{j,k} = \sum_{\xi} \ket{\Psi_{(\beta, \mathfrak{t}^{\beta}_{\xi}, j)}}\bra{\Psi_{(\beta, \mathfrak{t}^{\beta}_{\xi}, k)}}.
\ee
Thanks to our previous choice of taking the basis in such a way that the AKLT states are precisely basis vectors, then some of the projectors defined in Eq.~\eqref{equ::proj_1} correspond exactly to the MPS projectors given in Eq.~\eqref{equ::mps_proj}. In order to write the most general Hamiltonian on $l$ sites one must consider all the other projectors, namely the ones which project onto a subsector of $\mathcal{Q}$, and the associated intertwiners,
\be
h = \sum_{\alpha\in\yset{Q}} \sum_{i=1}^{\mu^{\alpha}_{\mathcal{Q}}} \left( c^{\alpha}_{i,i} \mathbb{P}_{Q^{\alpha}_i} + \sum_{j\neq i}^{\mu^{\alpha}_{\mathcal{Q}}} c^{\alpha}_{i,j} \mathbb{X}^{\alpha}_{i,j} \right)
\ee
where $c^{\alpha}$ is a $\mu^{\alpha}_{\mathcal{Q}} \times \mu^{\alpha}_{\mathcal{Q}}$ symmetric positive definite matrix. Taking $c^{\alpha}$ to be the identity matrix, $c^{\alpha} = \mathds{1}_{\mu^{\alpha}_{\mathcal{Q}}}$ one obtains $h = \mathbb{P}_{\mathcal{Q}}$, the projector onto the entire subspace $\mathcal{Q}$, namely one recovers the MPS Hamiltonian.

To summarize, the MPS method allows one to build a projector onto the space $\mathcal{K}$, which becomes the kernel of the Hamiltonian by defining
\be
h_{\MPS} = \mathds{1} - \mathbb{P}_{\MPS} \equiv \mathds{1} - \mathbb{P}_{\mathcal{K}}.
\ee
Alternatively one can build a parent Hamiltonian by projecting onto $\mathcal{Q}$ and one has
\be
h = \mathbb{P}_{\mathcal{Q}} =  \mathds{1} - \mathbb{P}_{\mathcal{K}} = h_{\MPS}.
\ee
But the decomposition of $\mathcal{Q}$ into different sectors with definite $\SU{N}$ quantum numbers allows us to extend this definition and to project separately on each subsector, with a different positive coefficient. Moreover transitions (swaps) within subsectors having the same $\SU{N}$ quantum numbers are also permitted and are realized by the action of the intertwiners.

\subsection{Examples: Spin-$1$ AKLT Hamiltonians}

\begin{figure}
\begin{center}
\includegraphics[width=0.96\textwidth]{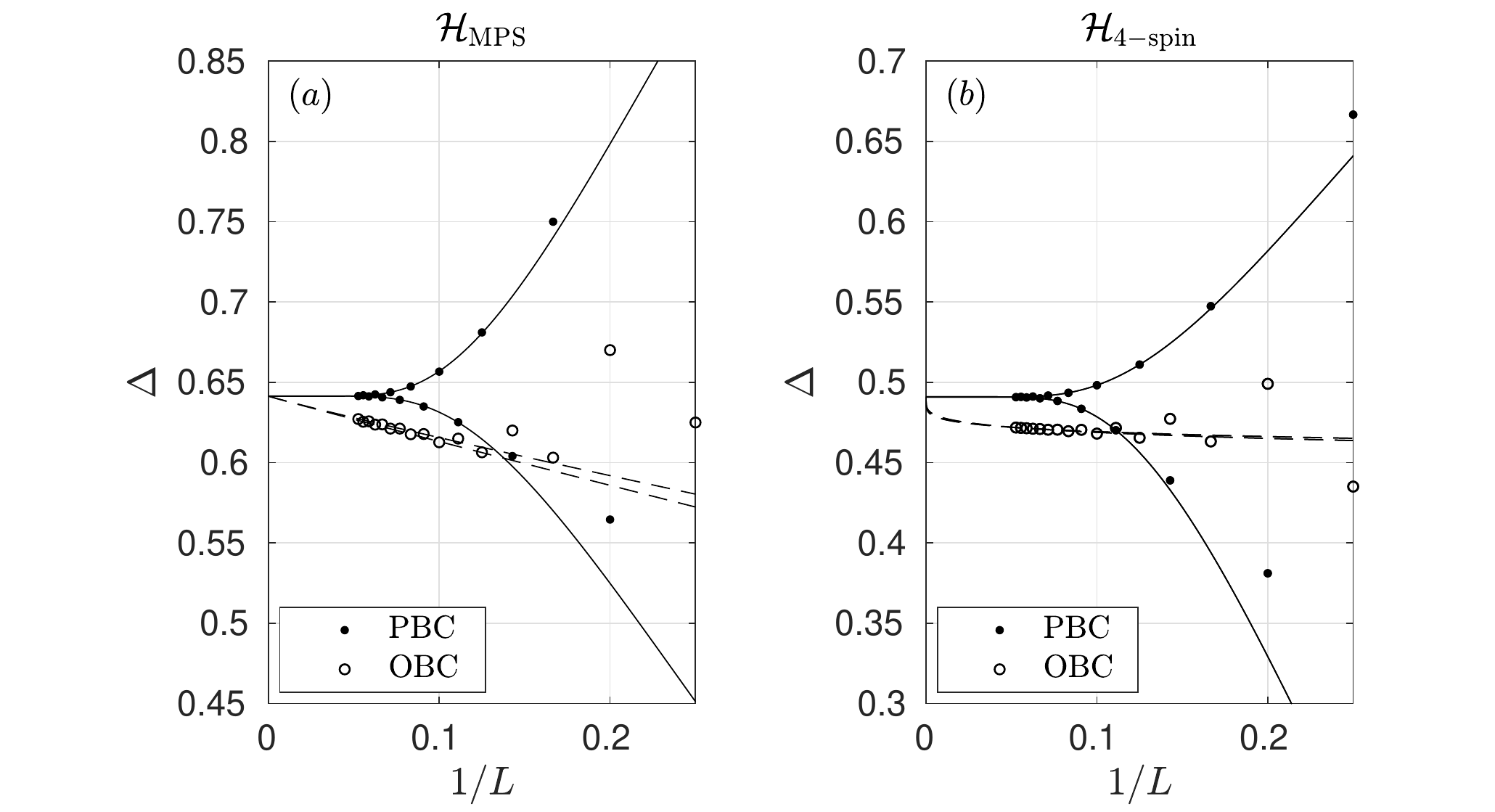}
\end{center}
\caption{Bulk gap $\Delta$ of the spin-$1$ AKLT Hamiltonian with spin-$1$ edge states constructed (a) from the MPS wave function and (b) from the operators acting in the subspace $\mathcal{Q}$. Full lines are exponential fits w.r.t $1/L$ while dashed lines are power-law fits. One obtains $\Delta_{\MPS} \simeq 0.641$ and $\Delta_{4{\rm-spin}} \simeq 0.491$.}
\label{fig::su2::pS1vS1::gap}
\end{figure}

The spin-$1$ AKLT state with spin-$1$ edge states can again be used as an illustrative example for the construction of parent Hamiltonians. Proceeding as explained above from the MPS wave function on $3$ sites we obtain a $3$-site local Hamiltonian $h_{\MPS} = \mathds{1} - (\mathbb{P}^{S=1}_{\MPS} + \mathbb{P}^{S=2}_{\MPS})$. This Hamiltonian being real, hermitian and reflection symmetric,  can be expanded in terms of at most $8$ SU($2$) invariant operators acting on $3$ sites (see Appendix~\ref{app:su2spin1}). We end up with
\be
\mathcal{H}_{\MPS} = \sum_i \tau_i(h_{\MPS})
\ee
where
\be
\begin{aligned}
h_{\MPS} = & \ 2 - \frac{1}{4} \bold{S}_{1} \cdot \bold{S}_{3} - \frac{1}{4} (\bold{S}_{1} \cdot \bold{S}_{3})^2  
-  \frac{5}{8} \left( (\bold{S}_{1} \cdot \bold{S}_{2})^2 + (\bold{S}_{2} \cdot \bold{S}_{3})^2 \right) \\
+ & \ \frac{3}{8} \left( (\bold{S}_{1} \cdot \bold{S}_{2}) (\bold{S}_{1} \cdot \bold{S}_{3}) (\bold{S}_{2} \cdot \bold{S}_{3}) + \text{h.c.} \right).
\end{aligned}
\label{equ::su2pS1::h_mps}
\ee
On a periodic chain of length $L$ the Hamiltonian reads
\be
\begin{aligned}
\mathcal{H}_{\MPS} = & \ \sum_{i=1}^L \bigg[ - \frac{5}{4} (\bold{S}_i \cdot \bold{S}_{i+1})^2 - \frac{1}{4} \bold{S}_i \cdot \bold{S}_{i+2} - \frac{1}{4} (\bold{S}_i \cdot \bold{S}_{i+2})^2 + 2\bigg] \\
+ & \ \frac{3}{8} \sum_{i=1}^L \bigg( (\bold{S}_i \cdot \bold{S}_{i+1}) (\bold{S}_{i} \cdot \bold{S}_{i+2}) (\bold{S}_{i+1} \cdot \bold{S}_{i+2}) + \text{h.c} \bigg).
\end{aligned}
\label{equ::su2pS1::h_mps_pbc}
\ee
The extrapolation of the bulk gap of this Hamiltonian versus $1/L$ is given in Fig.~\ref{fig::su2::pS1vS1::gap}(a).

Let us define now the projectors onto the different spin subspaces of $\bold{3}^{\otimes 3}$ as $\mathbb{P}^S_{\nu}$ where $\nu=1,...,\mu_S$ and where we explicitly take $\mathbb{P}^{S=1}_{3} = \mathbb{P}^{S=1}_{\MPS}$ as well as $\mathbb{P}^{S=2}_{2} = \mathbb{P}^{S=2}_{\MPS}$ (notice that $\mathbb{P}^{S=0} = \mathbb{P}^{S=0}_{\MPS}$ can equivalently be obtained from the Casimir construction, \ref{app:su2spin1}). We also introduce an ``intertwiner'' $\mathbb{S}$ which exchanges the two spin-$1$ irreps living in $\mathcal{Q}$ (here $\mathbb{S} \equiv \mathbb{X}^{S=1}_{1,2} + \mathbb{X}^{S=1}_{2,1}$ is hermitian to simplify the notation). The exact expressions of these operators are given in~\ref{app:su2spin1}. The most general parent Hamiltonian $h$ acting on $3$ sites can then be written as,
\be
h = c^{S=1}_{1,1} \mathbb{P}^{S=1}_{1} + c^{S=1}_{2,2} \mathbb{P}^{S=1}_{2} + c^{S=1}_{1,2} \mathbb{S} + c^{S=2} \mathbb{P}^{S=2}_{1} + c^{S=3} \mathbb{P}^{S=3}
\label{equ::su2pS1::most_general_h}
\ee
where the coefficients $c^{S=2}>0$, $c^{S=3}>0$ and where the symmetric $2\times 2$ matrix $c^{S=1}$ with elements $c^{S=1}_{i,j}$ must be positive definite. This ensures that $h$ is a positive semi-definite map, whose kernel is precisely the AKLT states. Equation~\eqref{equ::su2pS1::most_general_h} together with Table~\ref{table::gH::coeffs_projectors} as well as the positivity conditions on the real coefficients provides an entire family of parent Hamiltonians for the spin-$1$ AKLT state with spin-$1$ edge states. Note that, generically, inversion symmetry is explicitly broken unless $c^{S=1}_{1,1}=c^{S=1}_{2,2}$
and $c^{S=1}_{1,2}=0$. The MPS parent Hamiltonian given by Eq.~\eqref{equ::su2pS1::h_mps} corresponds to taking $c^{S=1} = \mathds{1}$, $c^{S=2} = c^{S=3} = 1$, which is the only way to get a projector. The reflection symmetric version ($c^{S=1}_{1,2}=0$) of the Hamiltonian~\eqref{equ::su2pS1::most_general_h} has been derived in Ref.~\cite{tu_topologically_2009}, however without reexpressing it in terms of spin operators. An explicit expression has then been obtained in Ref.~\cite{wen-jia_su3_2016}, where $c^{S=1}_{1,2} = 0$ is still assumed.

The freedom in the parameters in Eq.~\eqref{equ::su2pS1::most_general_h} can be used to get simpler parent Hamiltonians. For instance, by a judicious choice of the coefficients $c^S$, one can derive a parent Hamiltonian with only $2$-spin and $4$-spin operators:
\be
\begin{aligned}
h_{4{\rm-spin}} = & \ \frac{7}{3} - \frac{1}{2} \left( \bold{S}_1 \cdot \bold{S}_2 + \bold{S}_2 \cdot \bold{S}_3 \right) + \frac{4}{3} \bold{S}_1 \cdot \bold{S}_3  \\
- & \ \frac{2}{3} \left( (\bold{S}_1 \cdot \bold{S}_2)^2 + (\bold{S}_2 \cdot \bold{S}_3)^2 \right) + \frac{1}{3} (\bold{S}_1 \cdot \bold{S}_3)^2 \\
- & \ \frac{1}{2} \left( (\bold{S}_1 \cdot \bold{S}_2) (\bold{S}_2 \cdot \bold{S}_3) + \text{h.c.} \right).
\end{aligned}
\ee
In this Hamiltonian, we have removed the $6$-spin operator from the MPS Hamiltonian in Eq.~\eqref{equ::su2pS1::h_mps} at the price of introducing $4$-spin operators. On a periodic chain of length $L$ the Hamiltonian reads,
\be
\begin{aligned}
\mathcal{H}_{4{\rm-spin}} = & \ \sum_{i=1}^L \bigg[ - \bold{S}_i \cdot \bold{S}_{i+1} + \frac{4}{3} \bold{S}_i \cdot \bold{S}_{i+2} - \frac{4}{3} (\bold{S}_i \cdot \bold{S}_{i+1})^2 + \frac{1}{3} (\bold{S}_i \cdot \bold{S}_{i+2})^2 + \frac{7}{3}\bigg] \\
- & \ \frac{1}{2} \sum_{i=1}^L \bigg( (\bold{S}_i \cdot \bold{S}_{i+1}) (\bold{S}_{i+1} \cdot \bold{S}_{i+2}) + \text{h.c} \bigg).
\end{aligned}
\label{equ::su2::pS1vS1::Hamiltonian_simple_PBC}
\ee
Figure~\ref{fig::su2::pS1vS1::gap} compares the bulk gap of this parent Hamiltonian to the one of the MPS Hamiltonian given in Eq.~\eqref{equ::su2pS1::h_mps_pbc}.

\section{Interpolation of spin-$1$ AKLT Hamiltonians}
\label{sec::interpolspin1}

The original AKLT Hamiltonian Eq.~\eqref{equ::H_AKLT_87} lies in the Haldane phase, which is a symmetry protected topological phase (SPT phase). In general the parent Hamiltonian of the spin-$1$ chain with arbitrary half-odd-integer virtual spin is expected to be in a SPT phase protected by a $\mathbb{Z}_2\times\mathbb{Z}_2$ symmetry
(set of two orthogonal SU(2) $\pi$-rotations), unlike the parent Hamiltonians with integer virtual spins~\cite{kennedy_hidden_1992}. The construction of parent Hamiltonians for spin-$1$ AKLT states with arbitrary edge states allows us to study the transition between protected and unprotected phases by interpolating between the exactly solvable AKLT points. 
On general grounds, a critical point with central charge $c = 1$ is expected between a $\mathbb{Z}_2\times\mathbb{Z}_2$ SPT phase and a trivial 
phase~\cite{tsui_phase_2017}.

\begin{figure}
\begin{center}
\includegraphics[width=1.0\textwidth]{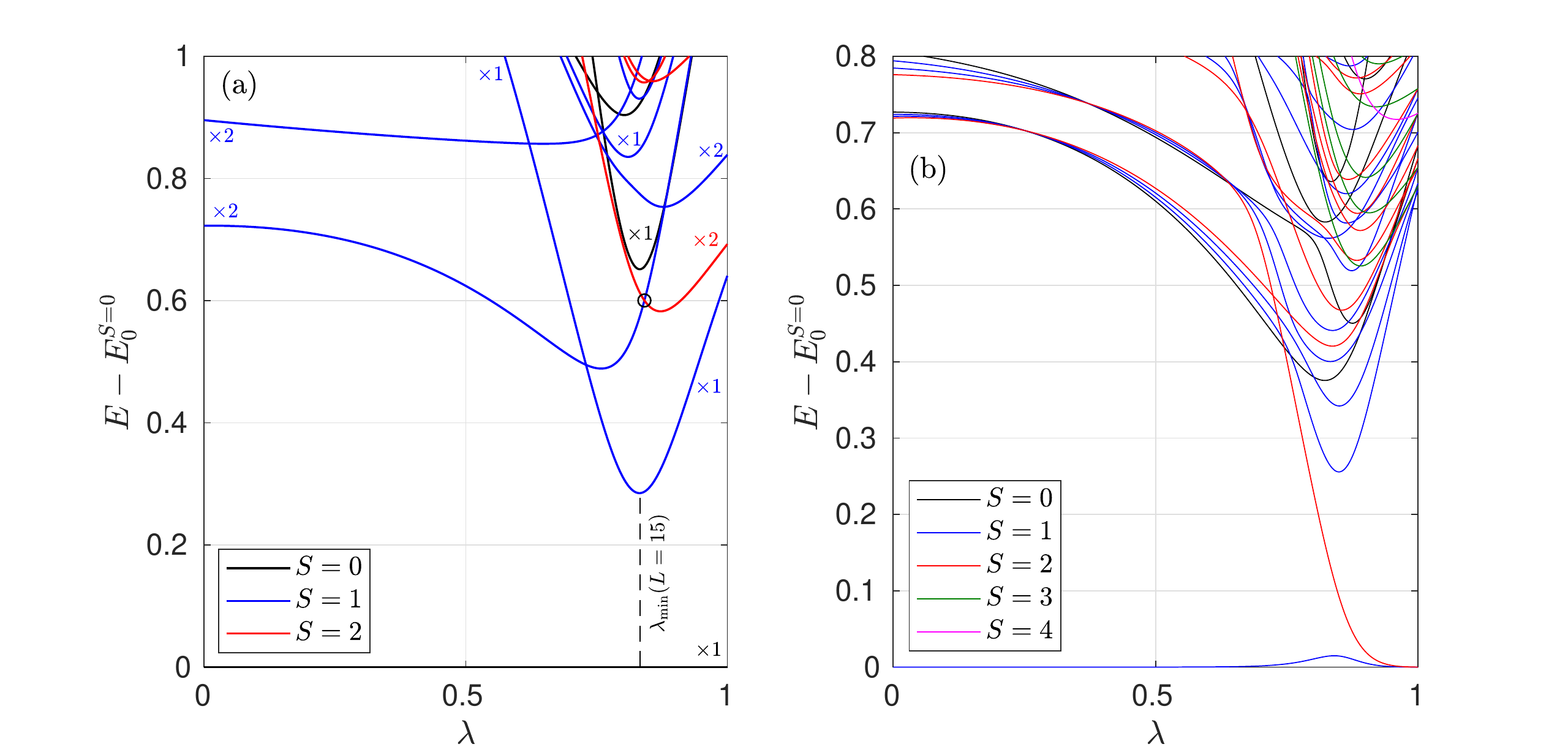}
\end{center}
\caption{Low energy spectrum of the Hamiltonian $\mathcal{H}(\lambda)$  (see text) on (a) periodic and (b) open chain of length $L=15$ versus $\lambda$. For PBC the minimum of the gap is indicated, as well as the position of the crossing of the second and third excited states. The extrapolation of the position of this crossing to infinite size does not lead to the approximate value of $\lambda_c$, preventing us to use level spectroscopy to extract more accurately the critical point. Degeneracies, corresponding to momenta $0, \pi$ or $\pm k$ are indicated in parenthesis. For OBC one sees a quintuplet of states coming from the bulk of $\mathcal{H}_{\AKLT}$ which collapses to zero energy at $\lambda=1$ and leads to a $9$-fold degeneracy of the zero-energy ground state.}
\label{fig::su2::spectrumPBCOBC}
\end{figure}

In order to study the transition we define the Hamiltonian
\be
\mathcal{H}(\lambda) = (1-\lambda) \mathcal{H}_{\AKLT} + \lambda \mathcal{H}_{\MPS}
\label{equ:interpol}
\ee
interpolating between the original AKLT spin-$1$ Hamiltonian  given in Eq.~\eqref{equ::H_AKLT_87} (of unique GS ${\ket{S=1, s=1/2}}$)
and the spin-$1$ Hamiltonian $\mathcal{H}_{\MPS}$ given in Eq.~\eqref{equ::su2pS1::h_mps_pbc} (of unique GS ${\ket{S=1, s=1}}$). The spectrum of this Hamiltonian is given in Fig.~\ref{fig::su2::spectrumPBCOBC}(a) for PBC and in Fig.~\ref{fig::su2::spectrumPBCOBC}(b) for OBC.

\begin{figure}
\begin{center}
\includegraphics[width=.96\textwidth]{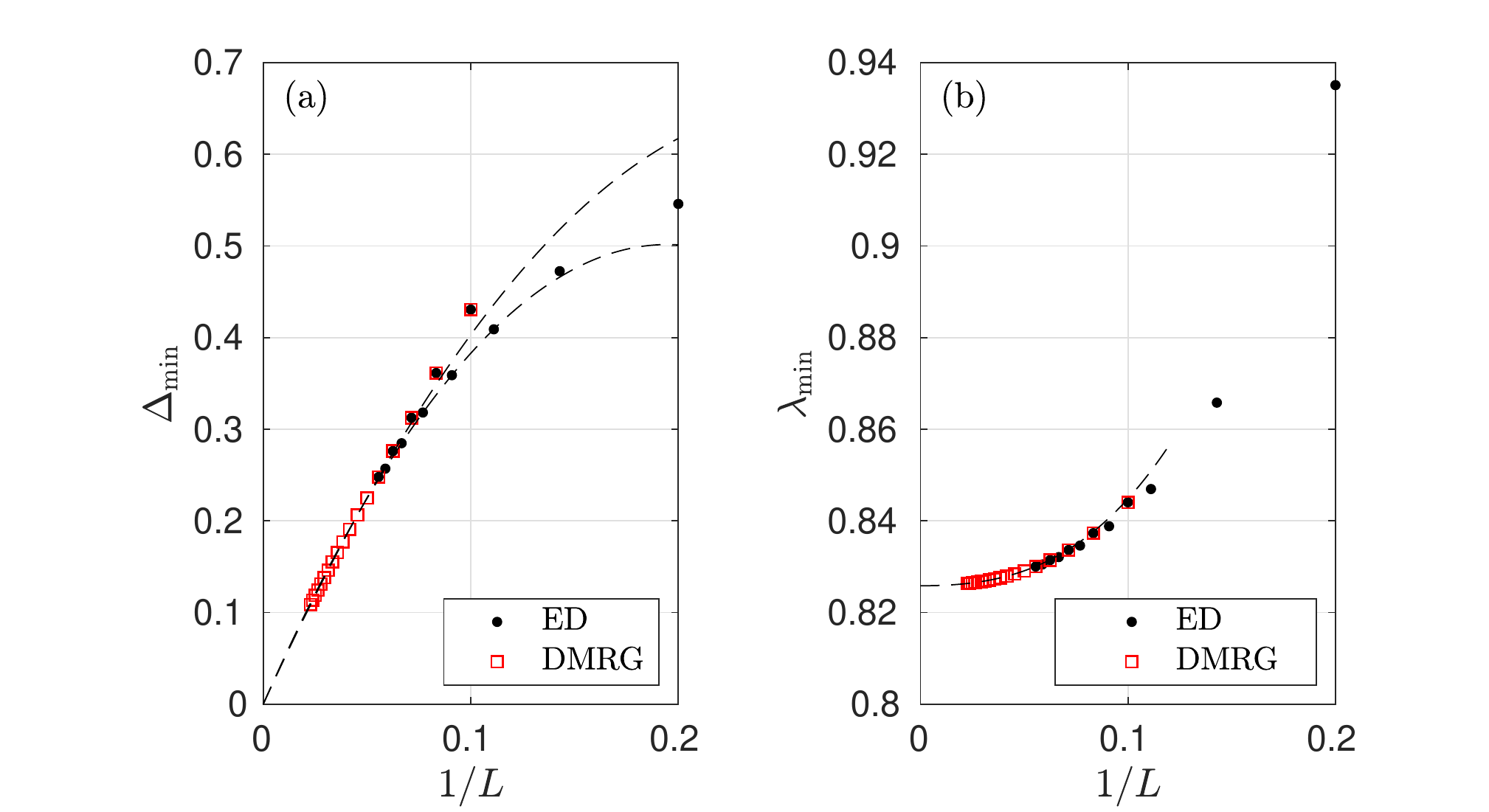}
\end{center}
\caption{(a) Minimum of the gap of the Hamiltonian $\mathcal{H}(\lambda)$ and (b) its position $\lambda_{\min}$, obtained with ED and DMRG. One gets $\lambda_c \simeq 0.8259(1)$. DMRG is for even lengths only ($L$ up to $44$ sites). In (b), the cubic fit with vanishing linear term is over even lengths $L \geqslant 20$ only.}
\label{fig::su2pS1::gapmin_lambdamin_v3}
\end{figure}

\begin{figure}
\begin{center}
\includegraphics[width=1.0\textwidth]{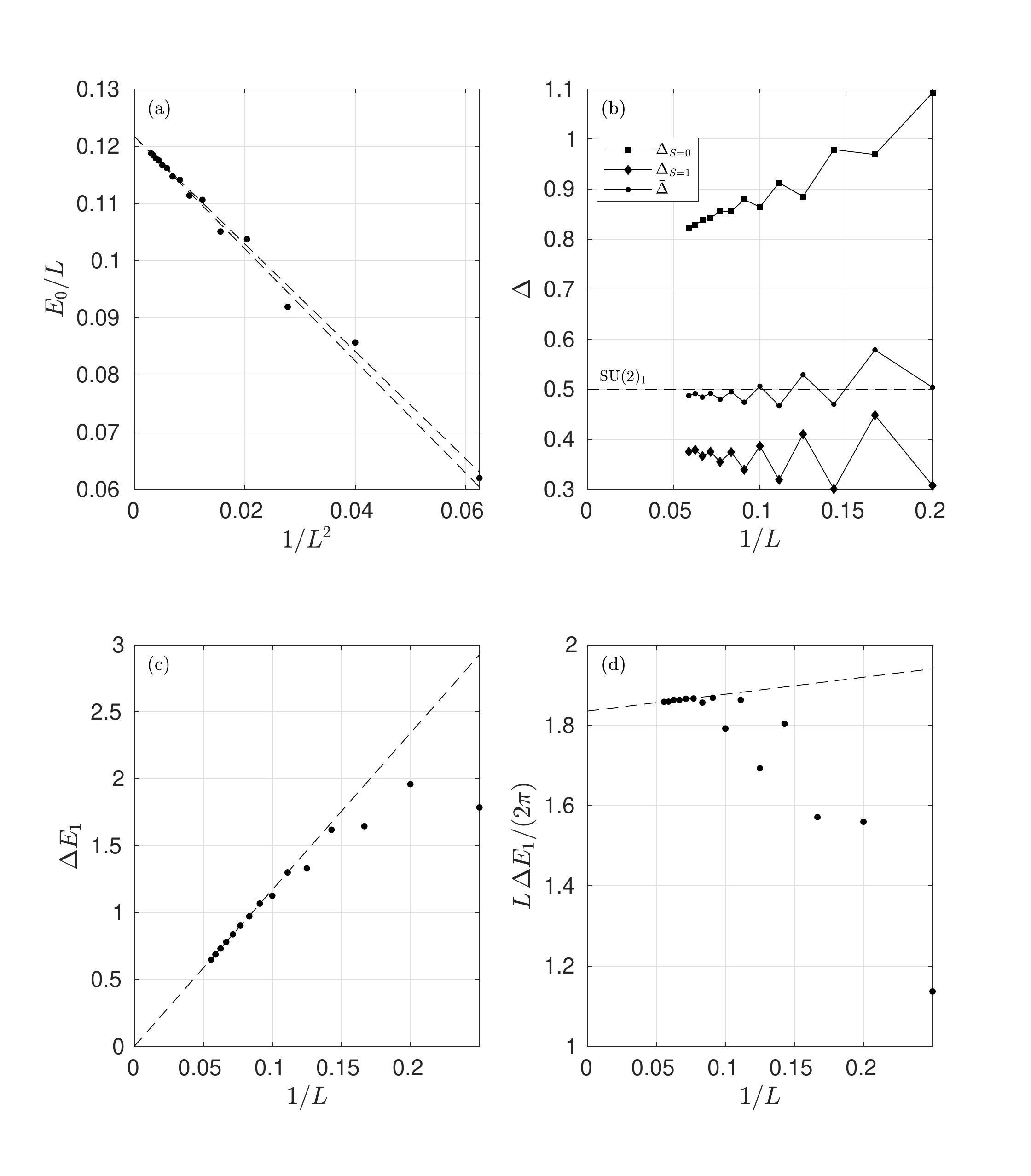}
\end{center}
\caption{Critical behavior of the periodic $L$-site chain at $\lambda_c = 0.8259$. (a) Ground state energy per site and fit to the CFT scaling formula. (b) Scaling dimensions of the primary fields associated to the first excited states. $\bar{\Delta} := (3 \Delta_{S=1} + \Delta_{S=0})/4$ is a combination which removes the logarithmic corrections. (c) Gap $\Delta E_1$ of the first excited state at momentum $2\pi/L$ and non-zero Casimir. (d) Velocity of light $v = L \Delta E_1/(2\pi)$.}
\label{fig::su2pS1::e0_velocity_lambda08259}
\end{figure}

We extract the position of the critial point $\lambda_c$ by extrapolating to infinite size the position of the mininum of the gap in the spectrum of $\mathcal{H}(\lambda)$. The results are obtained with a combination of exact diagonalization (ED) up to rings of size $L=18$~\cite{lin_exact_1990,nataf_exact_2016} and of DMRG up to rings of size $L=44$~\footnote{ITensor Library, \href{http://itensor.org}{http://itensor.org}.}. Figures~\ref{fig::su2pS1::gapmin_lambdamin_v3}(a) and~\ref{fig::su2pS1::gapmin_lambdamin_v3}(b) show the minimum of the gap and the associated values of $\lambda$ for the different system sizes. One obtains the value $\lambda_c := \lambda_{\min}(\infty) \simeq 0.8259$. At the critical point the ground state energy per site of the $L$-site chain is given by the CFT formula~\cite{blote_conformal_1986,affleck_universal_1986}
\be
\epsilon_0(L) = \frac{E_0(L)}{L} = \epsilon_0(\infty) - \frac{\pi c v}{6 L^2} + o(L^{-2})
\ee
where $\epsilon_0(\infty)$ is the energy per site in the thermodynamic limit, $c$ is the central charge and $v$ the velocity of light. The latter can be obtained from the energy $E_1(L)$ of the first excited state having momentum $2\pi/L$ and non-zero Casimir through the formula
\be
\Delta E_1(L) := E_1(L) - E_0(L) = \frac{2\pi v}{L} + o(L^{-1}).
\ee
In practice we observe that it is more accurate to fit $L \Delta E_1/(2\pi) = v + o(1)$ rather than simply $\Delta E_1$. The fits are shown in Fig.~\ref{fig::su2pS1::e0_velocity_lambda08259}. The factor $c v$ is averaged between its value for even and odd lengths (discrepancy of order $4$ percent). We finally extract the central charge $c \simeq 0.9995$, in very good agreement with the expected value $c = 1$. In order to explore further the conformal field theory at the transition point we plot in Fig.~\ref{fig::su2pS1::e0_velocity_lambda08259}(b) the scaling dimension of the primary fields associated to the first singlet and triplet excited states, and extract a combination which removes the logarithmic corrections~\cite{moreo_conformal_1987,ziman_are_1987}. The results seem to converge to $\Delta = 1/2$. This, together with the obtained central charge and the initial $\SU{2}$ symmetry of the spin model, strongly suggests that the phase transition is governed by the $\SU{2}_1$ WZW conformal field theory.

One can try to play the same game and investigate a linear interpolation between AKLT Hamiltonians with spin-$1$ and spin-$3/2$ edge states. It seems however that the transition is first order for the case we have looked at, as discussed in~\ref{app::su2_z2timesz2}.

Note that similar phase transitions between VBS phases have been observed in spin-$2$ chains. Denoting by $J_4$ the positive weight of the projector onto the spin-$4$ irrep on two neighboring sites, it has been shown that, when $J_4 = 1$, the transition between the VBS phase of the so-called ``Scalapino-Zhang-Hanke'' model with spin-$3/2$ edge states and that of the AKLT model with spin-$1$ edge states has central charge $c=2$ probably in the $\SU{2}_4$ WZW universality class~\cite{zheng_continuous_2011}, while for $J_4 \simeq 0.27$, there seems to be a multicritical point with central charge $c=5/2$ described by the ${\rm SO}(5)_1$ WZW conformal field theory~\cite{jiang_critical_2010,zang_topological_2010}.

\section{Application to SU($N$)}
\label{sec::SUN}

\subsection{General construction}

The novelty appearing for $\SU{N}$ with $N>2$ is the possibility to have a non-trivial outer multiplicity $\mu^{\phys}>1$ of an irrep $\phys$ in the tensor product of two virtual irreps. When the dimension of the virtual irrep increases some physical irreps appear with an increasing outer multiplicity. For instance one can form only one $\SU{3}$ AKLT state $\ket{\shsc{3pt}{0.4}{\ydiagram{4,2}}, \shsc{3pt}{0.4}{\ydiagram{2,1}}}$ but one can construct two AKLT states $\ket{\shsc{3pt}{0.4}{\ydiagram{4,2}}, \shsc{3pt}{0.4}{\ydiagram{4,2}},\nu}, \nu = 1,2$. Tables~\ref{table::su3::aklt_tensors} and~\ref{table::su4::aklt_tensors} give the number of AKLT states one can construct for a selection of virtual and physical irreps in $\SU{3}$ and $\SU{4}$, respectively. In what follows we will focus on the $3$ simplest $\SU{3}$ AKLT states with virtual space $\vir = \ \shsc{3pt}{0.4}{\ydiagram{2,1}}$ highlighted in the first column of Table~\ref{table::su3::aklt_tensors}. We will also consider two series of simple $\SU{N}$ AKLT states, $N$ even, with fully antisymmetric virtual spaces corresponding to the first column of Table~\ref{table::su4::aklt_tensors} for $N = 4$.

\subsection{$\SU{3}$ AKLT state with physical $3$-box symmetric irrep}

\begin{figure}
\begin{minipage}[c][30mm]{1.0\textwidth}
\input{fig_su3_adjkronadj}
\end{minipage}
\begin{minipage}[c][30mm]{1.0\textwidth}
\input{fig_su3_content_irreps}
\end{minipage}
\begin{minipage}[c][50mm]{1.0\textwidth}
\input{fig_su3_sets}
\end{minipage}
\caption{Decomposition of the physical and virtual spaces in the case of $\SU{3}$ for the construction of the AKLT state $\ket{\phys = \ \shsc{0pt}{0.4}{\ydiagram{3}}, \vir = \ \shsc{3pt}{0.4}{\ydiagram{2,1}} \, }$. (a) Tensor product representations appearing in $\phys\otimes\phys$ and $\vir\otimes\vir$. Here the space $\mathcal{U} \setminus \mathcal{K} \neq \emptyset$ corresponds to unphysical states which cannot be accessed on a $2$-site chain with physical irrep $\phys$. (b) Irrep content of the different subspaces. One notices in particular that the physical irrep $\phys \not\in \mathcal{K}$ but rather in $\mathcal{U}\setminus \mathcal{K}$, and similarly for the singlet irrep $\bullet$. (c) Schematic representation of the physical and virtual spaces.}
\label{fig::su3::new}
\end{figure}
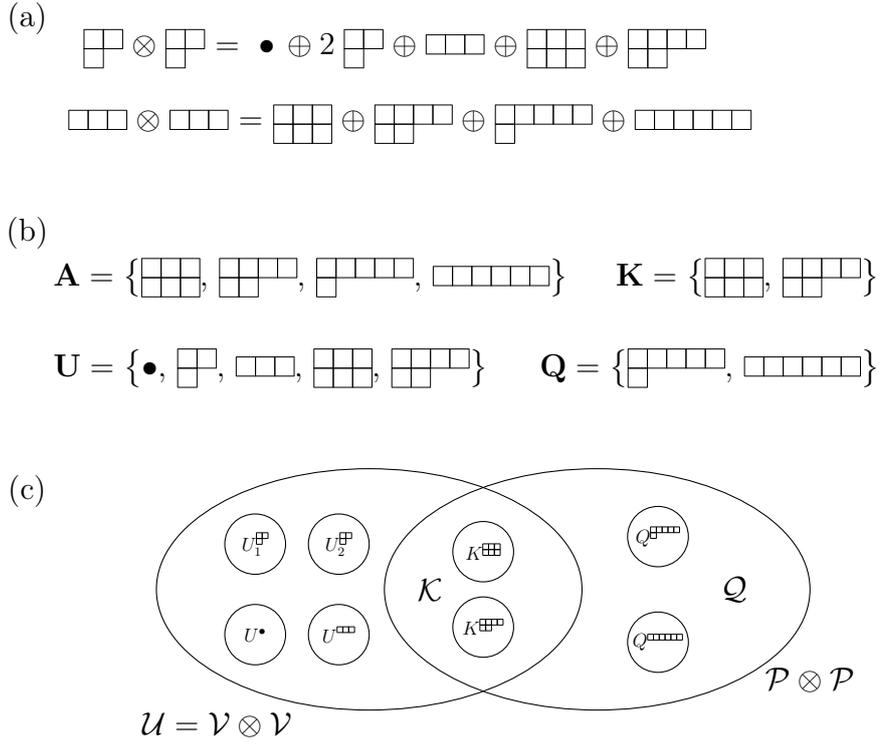

\begin{figure}
\begin{minipage}[t][2.5cm]{1.0\textwidth}
\input{fig_su3_aklt_epsilontensor.tex}
\end{minipage}
\caption{$\SU{3}$ AKLT state of Greiter and Rachel for the physical $3$-box symmetric irrep \protect\scalebox{0.4}{$\ydiagram{3}$} written in a MPS-like form. The ellipses denote a map from $3$ fundamental irreps to the symmetric irrep. This is easily obtained by writing the states of the irrep \protect\scalebox{0.4}{$\ydiagram{3}$} in terms of the virtual irreps with $3$ colors: $\ket{AAA}, \ket{BBB}, \ket{CCC}, ...$. The black filled circles denote antisymmetrizers which form extended singlet bonds on $3$ neighboring sites. Physical legs pointing upward or downward are equivalent.}
\label{fig::su3::p[300]GreiterRachel}
\end{figure}
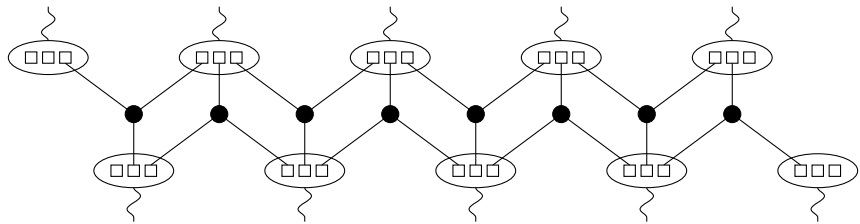

In Ref.~\cite{greiter_valence_2007} Greiter and Rachel introduced various new AKLT states and parent Hamiltonians for $\SU{N}$. A $\SU{3}$ VBS state with physical $3$-box symmetric irrep $\shsc{0pt}{0.4 }{\ydiagram{3}}$ is described in terms of fundamental representations: on each site, $3$ virtual fundamental irreps are projected onto the physical irrep, and extended singlets are formed on $3$ neighboring sites. A valid parent Hamiltonian could be derived using the quadratic Casimir operator: due to the formation of the singlets the state of two neighboring physical sites can be in any of the irreps appearing in
\be
\left( \shsc{0pt}{0.4}{\ydiagram{1}} \otimes \shsc{3pt}{0.4}{\ydiagram{1,1}} \, \right)^{\otimes 2} = 2\bullet \oplus \, 4 \shsc{3pt}{0.4}{\ydiagram{2,1}} \oplus \shsc{0pt}{0.4}{\ydiagram{3}} \oplus \shsc{3pt}{0.4}{\ydiagram{3,3}} \oplus \shsc{3pt}{0.4}{\ydiagram{4,2}}.
\ee
Using the result of the tensor product of two physical sites given in Fig.~\ref{fig::su3::new}(a) the simplest parent Hamiltonian is given by
\be
h = \mathds{1} - \left( \mathbb{P}^{\, \shsc{2pt}{0.15}{\ydiagram{3,3}}} + \mathbb{P}^{\, \shsc{2pt}{0.15}{\ydiagram{4,2}}}\right).
\label{equ::su3_h_GR}
\ee
By construction the above VBS is an AKLT state. However neither the MPS form of the state nor the exact nature of the edge states emerge from the construction. One can nevertheless provide an MPS-like picture for this state by using the fact that the formation of singlets on $3$ neighboring sites involves the full anti-symmetrization of $3$ fundamental irreps. This is obtained simply by the action of the L\'evi-Civita tensor. On the other hand the projection onto the physical irrep must keep only the $10$ fully symmetric states which are easily constructed with $3$ colors $A, B, C$: $\ket{AAA}, \ket{BBB}, \ket{CCC}, ...$. Figure~\ref{fig::su3::p[300]GreiterRachel} presents the AKLT state of Greiter and Rachel in MPS-like form. A proper MPS can be obtained by contracting the L\'evi-Civita tensor to the tensor located above or below. This leads to a MPS with auxiliary bond dimension $9$ and correlation length $\xi = 1/\ln 5$.

We revisit now the construction of this AKLT state in the framework of this paper. We define the AKLT state $\ket{\shsc{0pt}{0.4}{\ydiagram{3}}, \shsc{3pt}{0.4}{\ydiagram{2,1}}}$ which, by construction, has edge modes defined by the adjoint representation of $\SU{3}$ and auxiliary bond dimension $8$ (see Fig.~\ref{fig::su3::new}). The correlation length is given by $\xi=1/\ln 5$. The spectrum of the transfer matrix is actually precisely the same as the one of Greiter and Rachel except the absence of the additional $17$-fold degenerate eigenvalue $0$. Moreover the spectrum of the reduced density matrix coincide, except for the additional eigenvalue $0$ occuring in Greiter and Rachel's construction. Last but not least the overlap of the states with PBC is unity, showing unambiguously that the two constructions lead to the same AKLT state. In conclusion we have found an optimal representation of the AKLT state introduced by Greiter and Rachel for which the nature of the edge states is now manifest. We complement this claim with Table~\ref{table::su3_ed_GR} which shows the degeneracy of $0$-energy states of the Hamiltonian~\eqref{equ::su3_h_GR} derived by Greiter and Rachel (Eq.~(55) of Ref.~\citep{greiter_valence_2007}) in the relevant $\SU{3}$ subsectors~\cite{nataf_exact_2016}.

\begin{table}[b]
\caption{Degeneracy of the zero-energy spectrum of the Hamiltonian derived by Greiter and Rachel for the $\SU{3}$ AKLT state with physical $3$-box symmetric irrep $\protect\scalebox{0.4}{\ydiagram{3}}$. For OBC the zero-energy states belong precisely to the symmetry sectors defined by the irreps in $\shsc{3pt}{0.4}{\ydiagram{2,1}} \otimes \shsc{3pt}{0.4}{\ydiagram{2,1}}$. For PBC there is a unique singlet GS.}
\label{table::su3_ed_GR}
\begin{center}
\begin{tabular}{M{12mm}|M{12mm}M{12mm}M{12mm}M{12mm}M{12mm}N}
& $\bullet$ & $\shsc{2pt}{0.4}{\ydiagram{2,1}}$ & $\shsc{0pt}{0.4}{\ydiagram{3}}$ & $\shsc{2pt}{0.4}{\ydiagram{3,3}}$ & $\shsc{2pt}{0.4}{\ydiagram{4,2}}$ &\\[10pt]
\hline
PBC & $1$ & $0$ & $0$ & $0$ & $0$&\\[8pt]
OBC & $1$ & $2$ & $1$ & $1$ & $1$&\\[8pt]
\end{tabular}
\end{center}
\end{table}

\subsection{$\SU{3}$ AKLT state with physical adjoint irrep}
\label{sec::su3::phys_adj}

\begin{figure}
\begin{center}
\includegraphics[width=0.8\textwidth]{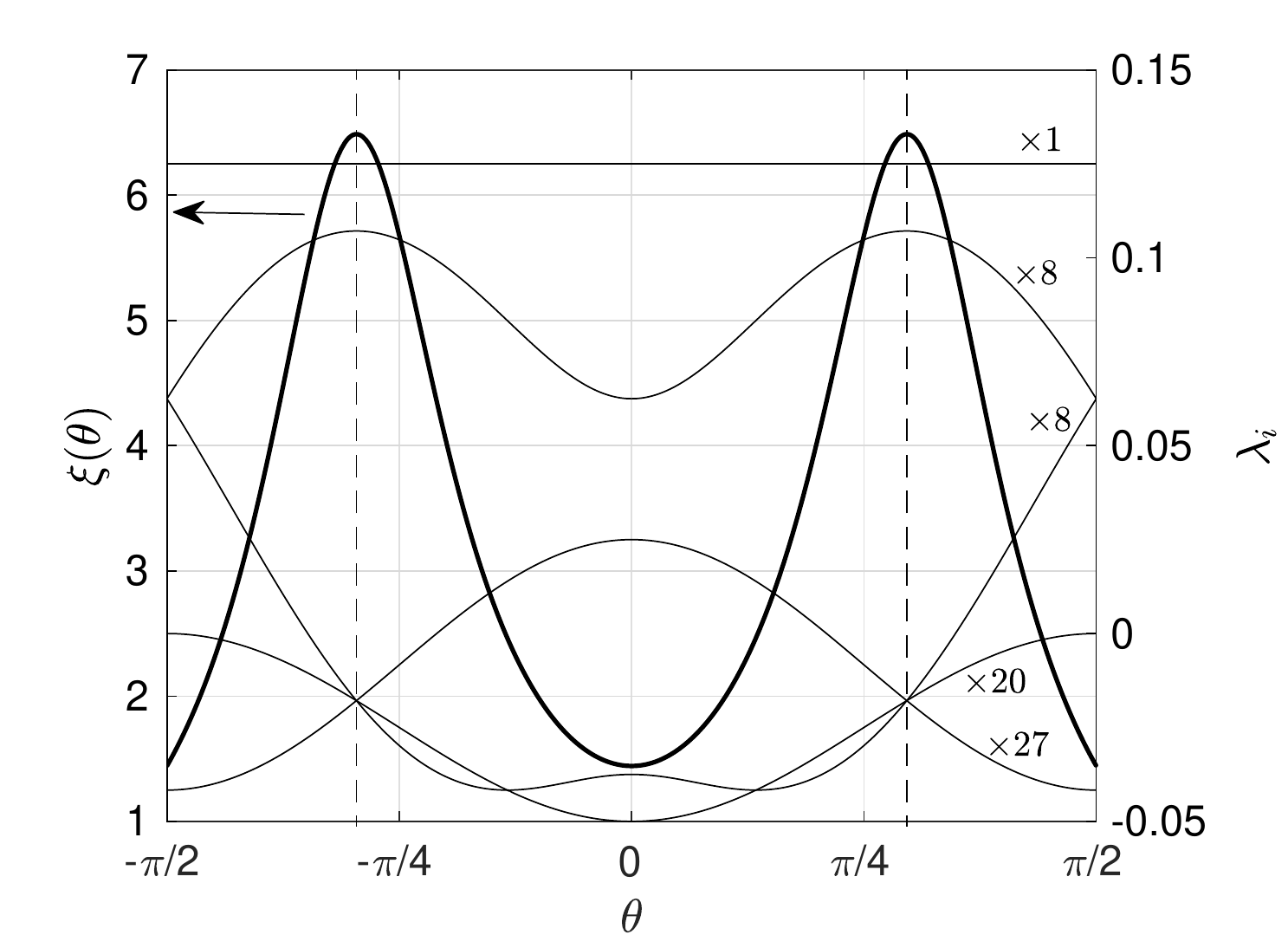}
\end{center}
\caption{Correlation length $\xi$ (thick line) and eigenvalues $\lambda_i$ of the transfer matrix versus angle $\theta$ of the $\SU{3}$ AKLT state $\ket{\shsc{3pt}{0.4}{\ydiagram{2,1}}, \shsc{3pt}{0.4}{\ydiagram{2,1}}, \theta}$. The spectrum of the transfer matrix is degenerate and the multiplicities correspond precisely to the dimensions of the irreps appearing in $\vir \otimes \vir$.}
\label{fig::su3::xi_and_lambdai}
\end{figure}

A natural and interesting extension of the previous section consists in taking the adjoint irrep of $\SU{3}$ to be both the virtual and the physical spin. This corresponds also to an extension of our spin-$1$ AKLT state with spin-$1$ edge states since $S=1$ is the adjoint irrep of $\SU{2}$. Here, the crucial novelty is the two-fold multiplicity of the physical irrep in the tensor product of two virtual irreps (see Fig.~\ref{fig::su3::p[300]::v[210]}(a)), which leads to two different AKLT states. In order to characterize these states we notice that the CGC associated with $\shsc{3pt}{0.4}{\ydiagram{2,1}} \otimes \shsc{3pt}{0.4}{\ydiagram{2,1}} \ \rightarrow \ \shsc{3pt}{0.4}{\ydiagram{2,1}}$ can be chosen to be either symmetric or antisymmetric under the exchange of the virtual spins~\cite{roy_chiral_2018}. One thus has a symmetric (under reflection) AKLT state $\ket{\shsc{3pt}{0.4}{\ydiagram{2,1}}, \shsc{3pt}{0.4}{\ydiagram{2,1}}, +}$ and an antisymmetric AKLT state $\ket{\shsc{3pt}{0.4}{\ydiagram{2,1}}, \shsc{3pt}{0.4}{\ydiagram{2,1}}, -}$ (we replaced the multiplicity index $\nu=1,2$ by $\nu=+,-$). These two states have correlation length $\xi_+ = \xi_- = 1/\ln 2$ and, for each of them, we are able to construct a $3$-site local, gapped, reflection symmetric and $\SU{3}$-invariant parent Hamiltonian (see Sec.~\ref{sec::su3_parent_H}). One can also build AKLT states which do not have a well-defined symmetry under reflection by mixing the CGC of the symmetric and antisymmetric tensors. Defining
\be
M^{\sigma}_{a,b}(\theta) = \cos(\theta) M^{\sigma,+}_{a,b} + \sin(\theta) M^{\sigma,-}_{a,b}
\label{equ::su3_combination_tensors}
\ee
one has access to a continuous family of AKLT states parametrized by the angle $\theta$ and denoted by $\ket{\shsc{3pt}{0.4}{\ydiagram{2,1}}, \shsc{3pt}{0.4}{\ydiagram{2,1}}, \theta}$. Except at $\theta = 0, \pi/2$ the state $\ket{\shsc{3pt}{0.4}{\ydiagram{2,1}}, \shsc{3pt}{0.4}{\ydiagram{2,1}}, \theta}$ is neither even nor odd under reflection. Figure~\ref{fig::su3::xi_and_lambdai} presents the spectrum of the transfer matrix associated with these states and their correlation lengths versus the angle $\theta$. 
It is worth mentioning a remarkable fact: the transfer matrix has a very structured spectrum. Its eigenvalues are degenerate with multiplicities corresponding to the dimensions of the irreps appearing in the tensor product of two virtual irreps. Moreover the largest eigenvalue is always the non-degenerate one (dimension of the singlet irrep). Furthermore we observed that the largest eigenvalue of the transfer matrix has the following expression,
\be
\lambda_{\text{max}} = \frac{\dim(\phys)}{\dim(\vir)^2}.
\label{equ::lambdamax_TM}
\ee
We have not been able to prove the degeneracy of the eigenvalues nor Eq.~\eqref{equ::lambdamax_TM}, but these claims are supported by all cases treated in this paper, and we have not been able to find any counterexample\footnote{These claims are also true for AKLT states constructed from non self-conjugate irreps. For instance the $\SU{N}$ AKLT states studied in Ref.~\cite{morimoto_z_2014,roy_chiral_2018} also have these properties. In that case analytical expressions for the eigenstates and eigenvalues of the transfer matrix and their degeneracies could be obtained thanks to the special properties of~\mbox{the}~\mbox{adjoint}~\mbox{irrep}}.

\subsection{$\SU{3}$ AKLT state with physical irrep $\phys = \ \shsc{2pt}{0.4}{\ydiagram{4,2}}$}

We discuss now briefly the last possible AKLT state with adjoint edge states (apart from the state $\ket{\shsc{3pt}{0.4}{\ydiagram{3,3}},\shsc{3pt}{0.4}{\ydiagram{2,1}}}$ which is similar to $\ket{\shsc{0pt}{0.4}{\ydiagram{3}},\shsc{3pt}{0.4}{\ydiagram{2,1}}}$). This state has physical irrep $\phys = \ \shsc{2pt}{0.4}{\ydiagram{4,2}}$ of dimension $27$. One can build only one such state, which has correlation length $\xi = 1/\ln 3$. A different construction consists in taking two fundamental and two conjugate irreps on each physical site, projecting them on the physical irrep on each site, and simultaneously forming two singlets on each bond between neighboring sites, Fig.~\ref{fig::su3::420}.
\begin{figure}
\input{fig_su3_aklt_irrep420_33bar}
\caption{$\SU{3}$ AKLT state with irrep $\phys = \ \shsc{3pt}{0.4}{\ydiagram{4,2}}$ on each site, formed by projecting two fundamental and two conjugate irreps on each site onto the physical irrep and forming two singlet bonds between every pair of neighboring sites.}
\label{fig::su3::420}
\end{figure}
The resulting MPS has auxiliary bond dimension $9$ while our construction has $D = 8$. Moreover the nature of the edge states is not manifest (although the presence of a fundamental and a conjugate irrep at both ends of an open chain strongly suggests the emergence of edge states belonging to the adjoint irrep).

\subsection{$\SU{3}$ parent Hamiltonians}
\label{sec::su3_parent_H}

Once the AKLT states are defined the associated parent Hamiltonians can be constructed along the lines of Sec.~\ref{sec::aklt_mps}. In particular,
for  the physical adjoint irrep $\phys = \ \shsc{3pt}{0.4}{\ydiagram{2,1}}$ one can build two $3$-site local reflection and $\SU{3}$-invariant parent Hamiltonians which lie in a phase with no spontaneous reflection symmetry breaking and with a unique ground state being either even or odd under reflection (these are the two states $\ket{\shsc{3pt}{0.4}{\ydiagram{2,1}}, \shsc{3pt}{0.4}{\ydiagram{2,1}}, \pm}$ discussed in Sec.~\ref{sec::su3::phys_adj}). This should be contrasted to the reflection-symmetric AKLT Hamiltonian discussed in Ref.~\cite{morimoto_z_2014}, in which the reflection symmetry of the two ground states is clearly spontaneously broken, or the pure Heisenberg Hamiltonian, which was claimed to lie in the same phase~\cite{rachel_spontaneous_2010}. When the angle $\theta \neq 0, \pi/2$ in Eq.~\eqref{equ::su3_combination_tensors} then the AKLT state $\ket{\shsc{3pt}{0.4}{\ydiagram{2,1}}, \shsc{3pt}{0.4}{\ydiagram{2,1}}, \theta}$ breaks reflection symmetry and so does explicitly its associated parent Hamiltonian, but the ground state degeneracy is still the expected one, by construction: unique ground state for PBC and $\mathcal{E} = D^2 = 64$ zero-energy states for OBC.

\subsection{$\SU{N}$, $N$ even: antisymmetric $\vir$ with $N/2$ rows}

We turn now to the case of $\SU{N}$, $N$ even. The case $N = 4$ has already been discussed in the litterature~\cite{nonne_symmetry-protected_2013,phases_capponi_2016,ueda_symmetry_2018}
and a selection of $\SU{4}$ AKLT states is reported in Table~\ref{table::su4::aklt_tensors}. We focus here on two families of simple $\SU{N}$ AKLT states constructed from the same
virtual fully antisymmetric irrep with $N/2$ boxes. A representative of each of these families for $N = 4$ appears in the first column $\vir =\ \shsc{2pt}{0.4}{\ydiagram{1,1}}$ of Table~\ref{table::su4::aklt_tensors}.

For a physical $\SU{4}$ square irrep $\shsc{3pt}{0.4}{\ydiagram{2,2}}$ of dimension $20$ the simplest choice consists in taking virtual spins living in the antisymmetric irrep of dimension~$6$. The correlation length of this state is given by $\xi = 1/\ln 5$. The generalization to $\SU{N}$ (with $N$ even) corresponds to a physical irrep $[2, \, 2, \, ..., \, 2]$ with $N/2$ rows (and two columns), made from two antisymmetric virtual irreps with $N/2$ boxes each (see Fig.~\ref{fig::su4irreps}(a)). The dimensions of these irreps as functions of $N$ are given by,
\be
\dim(\phys) = \frac{1}{N+1}\begin{pmatrix}
N+1\\
N/2 \\
\end{pmatrix}^2, \qquad
\dim(\vir) = \begin{pmatrix}
N \\
N/2 \\
\end{pmatrix}
\ee
where $\begin{pmatrix} n \\ p \end{pmatrix}$ is the binomial coefficient. 
\begin{figure}
\begin{minipage}[c][22mm]{1.0\textwidth}
\input{fig_su4_irreps}
\end{minipage}
\caption{(a) Physical rectangular irrep $\phys$ and virtual antisymmetric irrep $\vir$ for the generalization to $\SU{N}$ ($N$ even) of the $\SU{4}$ AKLT state $\ket{\shsc{3pt}{0.4}{\ydiagram{2,2}}, \shsc{3pt}{0.4}{\ydiagram{1,1}}}$. (b) Physical adjoint irrep of $\SU{N}$ and virtual irrep for the generalization to $\SU{N}$ of the $\SU{4}$ AKLT state $\ket{\text{Ad}, \shsc{3pt}{0.4}{\ydiagram{1,1}}}$.}
\label{fig::su4irreps}
\end{figure}
From numerical diagonalization of the transfer matrix up to $N = 12$ we conjecture a general formula for the correlation length,
\be
\xi = \frac{1}{\ln(N+1)}.
\ee
When $N = 2$ this AKLT state corresponds precisely to the original AKLT state, and so does the correlation length. 

With two virtual antisymmetric irreps with $N/2$ boxes ($N$ even) one can also form AKLT states with physical adjoint irrep (see Fig.~\ref{fig::su4irreps}(b))~\footnote{The tensor product of two antisymmetric irreps with $N/2$ boxes each can be decomposed into the direct sum of all irreps with two columns, $p$ boxes in the second column and $N-p$ boxes in the first column, $0 \leqslant p\leqslant N/2$.}. In this case, again based on numerical exact diagonalization of the transfer matrix, we conjecture the following general form for the correlation length,
\be
\xi^{-1} = \ln\left(\frac{N+1}{N-3}\right).
\ee
Here the correlation length increases with $N$, by contrast to the case of the irrep with two columns, where the correlation length decreases with $N$.\\

\section{Conclusion}
\label{sec::conclusion}

In this paper, we have introduced a systematic construction of gapped $\SU{N}$-symmetric one-dimensional spin liquids exhibiting edge states in open chains. This is based on a straightforward extension of the original AKLT procedure applied to an $\SU{2}$ spin-$1$ chain. For this purpose we have used the MPS framework in which each on-site physical $\SU{N}$ spin $\phys$ (characterized by a given Young tableau) is split into two identical virtual $\SU{N}$ irreps $\vir$ (virtual spins). 
For such a procedure to be realizable one then needs that the fusion product of two (identical) virtual spins contains (i) the physical irrep (possibly with a multiplicity) and (ii) the $\SU{N}$ singlet. The condition (ii) is necessary to realize maximally entangled nearest neighbor singlet bonds from all pairs of neighboring virtual spins, before the on-site projections onto the physical states. Therefore, the virtual spin $\vir$ should be characterized by a {\it self-conjugate} irrep of $\SU{N}$. Moreover, any fusion output of $\vir^{\otimes 2}$, for any $\vir$ self-conjugate, 
is a potential candidate for a valid AKLT state, or several AKLT states if this fusion output appears with a non-trivial outer multiplicity. Following such a procedure, we have proposed a selection of classes of simple $\SU{3}$ and $\SU{4}$ AKLT states, as well as two remarkable series of $\SU{N}$ AKLT state for all $N$ even. 

The existence of edge modes in open chains follows directly from our AKLT construction: after cutting a (large) periodic ring between two sites, one is left with a single unpaired virtual state at each end. Generically, the virtual states remain confined at each end of the chain in a region set by the bulk correlation length. Such edge states are {\it topologically} protected whenever they cannot fuse with (bulk) physical degrees of freedom $\phys$ to give rise to a non-degenerate ground state of the open chain.  Such a process could e.g. be forbidden thanks to the discrete symmetry subgroup $\mathbb{Z}_N\times\mathbb{Z}_N$ of the global $\SU{N}$ symmetry group (SPT phases)~\cite{tsui_phase_2017,duivenvoorden_topological_2013,duivenvoorden_symmetry_2013}. In our case, this occurs if $N$ is even and if the number of boxes in the Young tableau defining $\vir$ is $N/2 \ {\rm mod}(N)$. Then, the edge states can only disappear if the correlation length diverges at a quantum phase transition (or if a discontinuous first-order transition occurs). We have provided explicit examples  of edge physics and phase transitions by constructing exact {\it local} parent Hamiltonians (some given by exact analytic expressions) of $\SU{2}$ and $\SU{3}$ AKLT states with different types of edge modes. In the case of $\SU{2}$ spin-$1$ chains, $3$-site and $4$-site parent Hamiltonians with spin-$1$ and spin-$3/2$ edge states, respectively, can be constructed. Using an interpolation, this enabled us to investigate the transitions between two spin-$1$ SPT phases (with half-integer $s = 1/2$ and $s = 3/2$ edge states) and the (unprotected) spin-$1$ phase with spin-$1$ edge states. It would be interesting to study the same kind of transitions for $\SU{N}$ models, in which case the phase transition can be expected to be in other universality classes~\cite{tsui_phase_2017}. This is technically slightly more difficult however because of the increasingly more complicated form of the parent Hamiltonians, and this is left for future investigation.

Finally, this kind of construction can easily be extended to VBS in higher dimensions. The virtual irreps should still be self-conjugate to build singlet bonds, but the physical irrep can be any irrep appearing in the product of $z$ virtual irreps, where $z$ is the coordination number. Work is in progress along these lines.
 
\section*{Acknowledgments}

We are grateful to Pierre Nataf, Norbert Schuch and Keisuke Totsuka for fruitful discussions. SG is especially grateful to Pierre Nataf for help on implementing exact diagonalizations with $\SU{N}$ symmetry. We are also grateful to Hong-Hao Tu for very insightful comments on the first version of this manuscript. The calculations have been performed using the facilities of the Scientific IT and Application Support Center of EPFL. 
This work has been supported by the Swiss National Science Foundation (SNF) and by the TNSTRONG  ANR-16-CE30-0025  and TNTOP ANR-18-CE30-0026-01 grants awarded by the French  Research  Council. IA is supported by NSERC of Canada Discovery Grant
04033-2016 and by the Canadian Institute for Advanced Research.


\appendix

\section{$\SU{2}$ invariant operators for spin-$1$ on 3 sites}
\label{app:su2spin1}

For an irrep $\phys$ of $\SU{N}$ on $l$ sites with the decomposition
\be
\left( \alpha^{\phys} \right)^{\otimes \, l} = \overset{p}{\underset{i=1}{\mathlarger{\mathlarger{\oplus}}}} \, \mu_i \,  \alpha_i
\label{equ::prod_irreps_appendix}
\ee
one can build $W$ $\SU{N}$ invariant operators acting on $l$ sites, where $W = W_1 + W_2$ and
\be
W_1 = \sum_{i=1}^{p} \mu_i, \qquad W_2 = \sum_{i=1}^p \mu_i (\mu_i - 1).
\ee
These operators can be chosen as follows: $W_1$ projection operators $\mathbb{P}^{\alpha_i}_{\nu}, i=1,...,p, \quad \nu=1,...,\mu_i$ onto the different irreps appearing in the decomposition~\eqref{equ::prod_irreps_appendix}; $W_2$ operators $\mathbb{X}^{\alpha_i}_{\nu,\bar{\nu}}, \ 1\leqslant\nu \neq \bar{\nu} \leqslant \mu_i$ which maps the $\nu$-th block of irrep $\alpha_i$ onto the $\bar{\nu}$-th block if $\alpha_i$ has a non-trivial outer multiplicity $\mu_i>1$ in Eq.~\eqref{equ::prod_irreps_appendix}. These operators can further be combined into $W_2/2$ time-reversal (TR) symmetric operators, $\mathbb{S}^{\alpha_i}_{(\nu,\bar{\nu})} = \mathbb{X}^{\alpha_i}_{\nu,\bar{\nu}} + \mathbb{X}^{\alpha_i}_{\bar{\nu},\nu}$ and $W_2/2$ TR-antisymmetric (or ``chiral'') operators $\mathbb{A}^{\alpha_i}_{[\nu,\bar{\nu}]} = i( \mathbb{X}^{\alpha_i}_{\nu,\bar{\nu}} - \mathbb{X}^{\alpha_i}_{\bar{\nu},\nu})$. We thus end up with $m$ TR-symmetric and $n$ TR-antisymmetric operators with $m$ and $n$ given by
\be
m = \sum_{i=1}^{p} \frac{\mu_i (\mu_i+1)}{2}, \qquad n = \sum_{i=1}^{p} \frac{\mu_i (\mu_i-1)}{2}.
\ee
If the lattice of $l$ sites has an additional point-group symmetry (in our case, the reflection symmetry R), one can further characterize the operators wrt this point-group symmetry. Here the operators will be denoted as ``R-even/R-odd'' operators if they are reflection symmetric or antisymmetric, respectively.

For instance, for three $S=1$ spins on $3$ sites with the decomposition given in Eq.~\eqref{equ::pS1vS1::tensor_prod_1} one can build $W_1 = 7$ projection operators onto the different spin sectors, $6$ interchange operators $\mathbb{X}^{S=1}_{\nu,\bar{\nu}}, \ 1 \leqslant \nu \neq \bar{\nu} \leqslant 3$ within the spin-$1$ subspace and $2$ interchange operators $\mathbb{X}^{S=2}_{1,2}, \, \mathbb{X}^{S=2}_{2,1}$ within the spin-$2$ subspace. After performing the (anti)symmetrization of the interchange operators to obtain $\mathbb{S}^S_{(\nu,\bar{\nu})}$ and $\mathbb{A}^S_{[\nu,\bar{\nu}]}$ for $S=1,2$ one can reexpress all operators in terms of a (non-orthonormal) basis of combinations of spin operators which we chose as follows: $11$ TR-symmetric, hermitian and purely real operators, among them $8$ R-even and $3$ R-odd operators ($\mathcal{O}^3, \, \mathcal{O}^6$ and $\mathcal{O}^{11}$ are the R-odd operators):

\be
\begin{array}{lll}
\mathcal{O}^1 & = & \mathds{1}, \\[6pt]
\mathcal{O}^{2,3} & = & \frac{1}{2} \left( \bold{S}_1 \cdot \bold{S}_2 \pm \bold{S}_2 \cdot \bold{S}_3 \right), \\[6pt]
\mathcal{O}^{4} & = & \bold{S}_1 \cdot \bold{S}_3 \\[6pt]
\mathcal{O}^{5,6} & = & \frac{1}{2} \left( (\bold{S}_1 \cdot \bold{S}_2)^2 \pm (\bold{S}_2 \cdot \bold{S}_3)^2 \right),\\[6pt]
\mathcal{O}^{7} & = & (\bold{S}_1 \cdot \bold{S}_3)^2, \\[6pt]
\mathcal{O}^{8} & = & (\bold{S}_1 \cdot \bold{S}_2) (\bold{S}_1 \cdot \bold{S}_3) (\bold{S}_2 \cdot \bold{S}_3) + \text{h.c.}, \\[6pt]
\mathcal{O}^{9} & = & (\bold{S}_1 \cdot \bold{S}_2) (\bold{S}_2 \cdot \bold{S}_3) + \text{h.c.}, \\[6pt]
\mathcal{O}^{10,11} & = & \left( (\bold{S}_1 \cdot \bold{S}_3) (\bold{S}_2 \cdot \bold{S}_3) \pm (\bold{S}_1 \cdot \bold{S}_2) (\bold{S}_1 \cdot \bold{S}_3) \right) +  \text{h.c.},
\end{array}
\label{equ::su2spin1operators}
\ee

\noindent
and $4$ ``chiral'' ($3$ R-odd and $1$ R-even), hermitian, purely imaginary operators ($\mathcal{C}^4$ is the only R-even operator):
\be
\begin{array}{lll}
\mathcal{C}^1 & = & \bold{S}_1 \cdot \left( \bold{S}_2 \times \bold{S}_3 \right), \\[6pt]
\mathcal{C}^2 & = & i \left[ (\bold{S}_2 \cdot \bold{S}_3) (\bold{S}_1 \cdot \bold{S}_3) (\bold{S}_1 \cdot \bold{S}_2) - \text{h.c.}\right], \\[6pt]
\mathcal{C}^{3,4} & = & i \left[ \left( (\bold{S}_1 \cdot \bold{S}_2) (\bold{S}_2 \cdot \bold{S}_3) (\bold{S}_1 \cdot \bold{S}_3) \mp (\bold{S}_2 \cdot \bold{S}_3) (\bold{S}_1 \cdot \bold{S}_2) (\bold{S}_1 \cdot \bold{S}_3) \right) - \text{h.c.}\right].
\end{array}
\ee

The projectors onto the entire subspace of definite spin on $3$ sites can be obtained from the Casimir construction. Denoting $\bold{S}_T = \bold{S}_1 + \bold{S}_2 + \bold{S}_3$ the total spin we have,
\bea
\mathbb{P}^{S=0} & = & \frac{1}{288} \left( \bold{S}_T^2 - 2 \right)^2 \left( \bold{S}_T^2 - 6 \right) \left( \bold{S}_T^2 - 12 \right),\\
\mathbb{P}^{S=1} & = & \frac{1}{80} \bold{S}_T^2 \left( \bold{S}_T^2 - 6 \right) \left( \bold{S}_T^2 - 12 \right),\\
\mathbb{P}^{S=2} & = & \frac{1}{864} \bold{S}_T^2 \left( \bold{S}_T^2 - 2 \right) \left( \bold{S}_T^2 - 12 \right)^2,\\
\mathbb{P}^{S=3} & = & \frac{1}{720} \bold{S}_T^2 \left( \bold{S}_T^2 - 2 \right) \left( \bold{S}_T^2 - 6 \right) \label{equ::su2::pS1::projector_spin3}
\eea
and we recall that for spin-$1$ we have
\be
(\bold{S}_1 \cdot \bold{S}_2)^3 = 2 + \bold{S}_1 \cdot \bold{S}_2 - 2 (\bold{S}_1 \cdot \bold{S}_2)^2.
\ee
Choosing explicitly $\mathbb{P}^{S=1}_3 = \mathbb{P}^{S=1}_{\MPS}$ and $\mathbb{P}^{S=2}_2 = \mathbb{P}^{S=2}_{\MPS}$ all these operators can be expressed in terms of the $\SU{2}$ invariant operators defined in Eq.~\eqref{equ::su2spin1operators}, Table~\ref{table::gH::coeffs_projectors}, where we have used $\mathbb{S} \equiv \mathbb{S}^{S=1}_{(1,2)}$ to simplify the notation.

\begin{table}[h]
\caption{Expansion coefficients of the projectors $\mathbb{P}^S_{\nu}$ and of the intertwiner $\mathbb{S}$ onto the set of $\SU{2}$ invariant operators $\mathcal{O}^A, \ A=1,...,11$ defined in Eq.~\eqref{equ::su2spin1operators}. The projectors $\mathbb{P}^{S=1}_3$ and $\mathbb{P}^{S=2}_2$ are equivalent to the MPS projectors $\mathbb{P}^{S=1}_{\MPS}$ and $\mathbb{P}^{S=2}_{\MPS}$, respectively.}
\label{table::gH::coeffs_projectors}
\begin{center}
\begin{tabular}{M{8mm}|M{11mm}|M{11mm}|M{11mm}|M{11mm}|M{11mm}|M{11mm}|M{11mm}|M{11mm}N}
& $\mathbb{P}^{S=0}$ & $\mathbb{P}^{S=1}_1$ & $\mathbb{P}^{S=1}_2$ & $\mathbb{P}^{S=1}_3$ & $\mathbb{S}$ & $\mathbb{P}^{S=2}_1$ & $\mathbb{P}^{S=2}_2$ & $\mathbb{P}^{S=3}$ &\\[18pt]
\hline
$\mathcal{O}^1$ & $-\frac{1}{3}$ & $\frac{5}{7}$ & $\frac{31}{35}$ & $-1$ & $-\frac{4}{7} \sqrt{\frac{2}{5}}$ & $\frac{1}{3}$ & $\frac{1}{3}$ & $\frac{1}{15}$ &\\[15pt]
$\mathcal{O}^2$ & $-\frac{1}{3}$ & $0$ & $0$ & $0$ & $0$ & $-\frac{1}{3}$ & $\frac{1}{3}$ & $\frac{1}{3}$ &\\[15pt]
$\mathcal{O}^3$ & $0$ & $0$ & $0$ & $0$ & $0$ & $0$ & $0$ & $0$ &\\[15pt]
$\mathcal{O}^4$ & $\frac{1}{6}$ & $-\frac{11}{28}$ & $-\frac{16}{35}$ & $\frac{1}{4}$ & $\frac{3}{7\sqrt{10}}$ & $\frac{1}{2}$ & $-\frac{1}{6}$ & $\frac{1}{10}$ &\\[15pt]
$\mathcal{O}^5$ & $\frac{1}{3}$ & $-\frac{17}{28}$ & $-\frac{19}{35}$ & $\frac{3}{4}$ & $-\frac{3}{7 \sqrt{10}}$ & $-\frac{1}{6}$ & $\frac{1}{6}$ & $\frac{1}{15}$ &\\[15pt]
$\mathcal{O}^6$ & $0$ & $-\frac{1}{7}$ & $\frac{1}{7}$ & $0$ & $\frac{3}{7\sqrt{10}}$ & $0$ & $0$ & $0$ &\\[15pt]
$\mathcal{O}^7$ & $\frac{1}{6}$ & $-\frac{3}{28}$ & $-\frac{12}{35}$ & $\frac{1}{4}$ & $\frac{11}{7\sqrt{10}}$ & $\frac{1}{6}$ & $-\frac{1}{6}$ & $\frac{1}{30}$ &\\[15pt]
$\mathcal{O}^8$ & $-\frac{1}{6}$ & $\frac{11}{56}$ & $\frac{8}{35}$ & $-\frac{1}{8}$ & $-\frac{3}{14 \sqrt{10}}$ & $-\frac{1}{12}$ & $-\frac{1}{12}$ & $\frac{1}{30}$ &\\[15pt]
$\mathcal{O}^9$ & $0$ & $\frac{1}{14}$ & $\frac{1}{35}$ & $0$ & $\frac{1}{7} \sqrt{\frac{2}{5}}$ & $-\frac{1}{6}$ & $0$ & $\frac{1}{15}$ &\\[15pt]
$\mathcal{O}^{10}$ & $\frac{1}{12}$ & $-\frac{1}{28}$ & $-\frac{1}{70}$ & $0$ & $-\frac{1}{7\sqrt{10}}$ & $0$ & $-\frac{1}{12}$ & $\frac{1}{20}$ &\\[15pt]
$\mathcal{O}^{11}$ & $0$ & $-\frac{1}{14}$ & $\frac{1}{14}$ & $0$ & $\frac{3}{14\sqrt{10}}$ & $0$ &  $0$ & $0$ &\\[15pt]
\end{tabular}
\end{center}
\end{table}

\section{Phase transition to $\mathbb{Z}_2\times\mathbb{Z}_2$ SPT phase}

\label{app::su2_z2timesz2}

\begin{figure}
\begin{center}
\includegraphics[width=.96\textwidth]{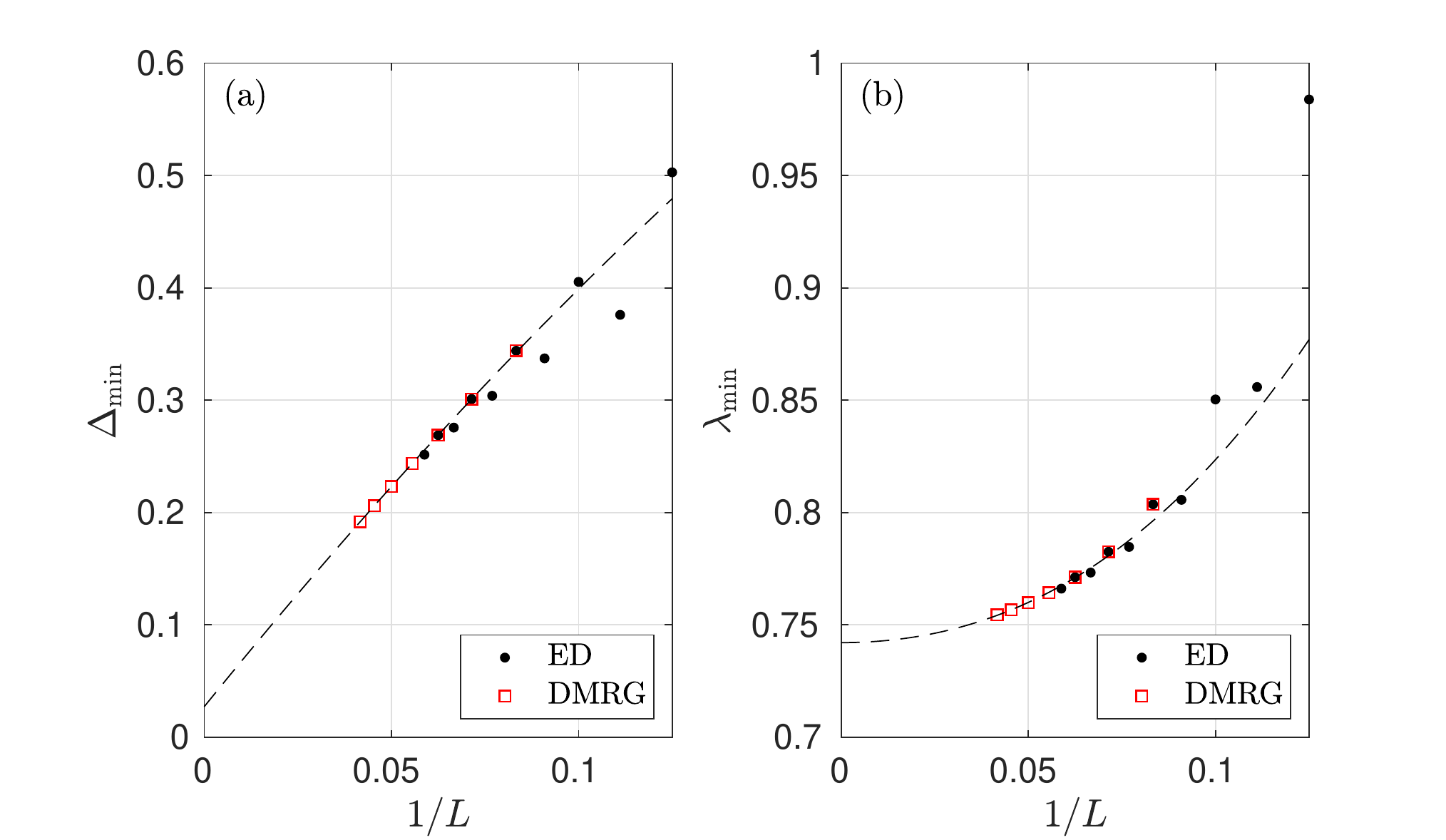}
\end{center}
\caption{(a) Minimum of the gap of the Hamiltonian $\mathcal{H}(\lambda)$ and (b) its position $\lambda_{\min}$ for the interpolation between $\ket{S=1, s=1}$ and $\ket{S=1, s=3/2}$.}
\label{fig::su2pS1::vS3half::gapmin_lambdamin}
\end{figure}

For the transition from $\ket{S=1, s=1}$ to $\ket{S=1, s=3/2}$, Figure~\ref{fig::su2pS1::vS3half::gapmin_lambdamin} shows the minimum of the gap and its position along the interpolation line between the MPS parent Hamiltonians of these two states. The results are in favor of a first order transition. A detailed study of the interpolation of these Hamiltonians along a different line possibly leading to a continuous phase transition is left for future work.\\

\section{Outer multiplicities for $\SU{3}$ and $\SU{4}$}

Tables~\ref{table::su3::aklt_tensors} and~\ref{table::su4::aklt_tensors} provide the outer multiplicity of a given physical irrep $\phys$ in the tensor product of two virtual irreps $\vir$, for a selection of self-conjugate virtual irreps $\vir$ of $\SU{3}$ and $\SU{4}$, respectively. These outer multiplicities thus correspond to the number of independent AKLT states $\ket{\phys, \vir, \nu}$ one can construct as described above.

\begin{table}[h]
\caption{Selection of $\SU{3}$ AKLT states $\ket{\phys,\vir}$ classified according to their physical and virtual irreps $\phys$ and $\vir$, displayed in the first column and first row, respectively. The table shows the number of independent  AKLT MPS (a cross means no AKLT state can be constructed). For all but the first virtual irrep there are physical irreps which are accessible but are not shown. The numbers in bold correspond to the 3 cases studied here.}
\label{table::su3::aklt_tensors}
\setlength{\tabcolsep}{3.7mm}
\begin{center}
\begin{tabular}{M{18mm}|M{16mm}M{16mm}M{16mm}M{16mm}N}
 & \multicolumn{4}{c}{$\shsc{-4pt}{1.0}{$\vir$}$} &\\[10pt]
$\phys$ & $\shsc{0pt}{0.45}{\ydiagram{2,1}}$ & $\shsc{0pt}{0.45}{\ydiagram{4,2}}$ & $\shsc{0pt}{0.45}{\ydiagram{6,3}}$ & $\left[ 8, 4 \right]$ &\\[20pt]
\cline{2-6}
$\shsc{6pt}{0.45}{\ydiagram{2,1}}$ & $\bold{2}$ & $2$ & $2$ & $2$ &\\[16pt]
$\shsc{6pt}{0.45}{\ydiagram{4,2}}$ & $\bold{1}$ & $3$ & $3$ & $3$ &\\[16pt]
$\shsc{6pt}{0.45}{\ydiagram{6,3}}$ & $\times$ & $2$ & $4$ & $4$ &\\[16pt]
$\left[ 8, 4 \right]$ & $\times$ & $1$ & $3$ & $5$ &\\[16pt]
$\left[ 10, 5 \right]$ & $\times$ & $\times$ & $2$ & $4$ &\\[16pt]
$\left[ 12, 6 \right]$ & $\times$ & $\times$ & $1$ & $3$ &\\[16pt]
$\shsc{0pt}{0.45}{\ydiagram{3}}$ & $\bold{1}$ & $1$ & $1$ & $1$ &\\[16pt]
$\shsc{0pt}{0.45}{\ydiagram{6}}$ & $\times$ & $1$ & $1$ & $1$ &\\[16pt]
$\left[ 9 \right]$ & $\times$ & $\times$ & $1$ & $1$ &\\[16pt]
$\left[ 12 \right]$ & $\times$ & $\times$ & $\times$ & $1$ &\\[16pt]
\end{tabular}
\end{center}
\end{table}

\begin{table}[h]
\caption{Selection of simple $\SU{4}$ AKLT states $\ket{\phys, \vir}$ classified according to their physical and virtual irreps $\phys$ and $\vir$, displayed in the first column and first row, respectively. The table shows the number of independent AKLT MPS (a cross means no AKLT  state can be constructed). For the $3$ last virtual irreps, there are additional physical irreps which are accessible but are not shown.
The numbers in bold correspond to the 2 cases studied in the text.}
\label{table::su4::aklt_tensors}
\begin{center}
\begin{tabular}{M{16mm}|M{12mm}M{12mm}M{12mm}M{12mm}M{12mm}M{12mm}N}
 & \multicolumn{6}{c}{$\shsc{-4pt}{1.0}{$\vir$}$} &\\[10pt]
$\phys$ & $\shsc{0pt}{0.45}{\ydiagram{1,1}}$ & $\shsc{0pt}{0.45}{\ydiagram{2,2}}$ & $\shsc{0pt}{0.45}{\ydiagram{3,3}}$ & $\shsc{0pt}{0.45}{\ydiagram{2,1,1}}$ & $\shsc{0pt}{0.45}{\ydiagram{4,2,2}}$ & $\left[ 6, 3, 3 \right]$ &\\[20pt]
\cline{2-7}
$\shsc{10pt}{0.45}{\ydiagram{2,1,1}}$ & $\bold{1}$ & $1$ & $1$ & $2$ & $2$ & $2$ &\\[16pt]
$\shsc{10pt}{0.45}{\ydiagram{4,2,2}}$ & $\times$ & $1$ & $1$ & $1$ & $3$ & $3$ &\\[16pt]
$\left[ 6, 3, 3 \right]$ & $\times$ & $\times$ & $1$ & $\times$ & $2$ & $4$ &\\[16pt]
$\left[ 8, 4, 4 \right]$ & $\times$ & $\times$ & $\times$ & $\times$ & $1$ & $3$ &\\[16pt]
$\left[ 10, 5, 5 \right]$ & $\times$ & $\times$ & $\times$ & $\times$ & $\times$ & $2$ &\\[15pt]
$\left[ 12, 6, 6 \right]$ & $\times$ & $\times$ & $\times$ & $\times$ & $\times$ & $1$ &\\[16pt]
$\shsc{6pt}{0.45}{\ydiagram{2,2}}$ & $\bold{1}$ & $1$ & $1$ & $1$ & $1$ & $1$ &\\[16pt]
$\shsc{6pt}{0.45}{\ydiagram{4,4}}$ & $\times$ & $1$ & $1$ & $\times$ & $1$ & $1$ &\\[16pt]
$\left[ 6, 6 \right]$ & $\times$ & $\times$ & $1$ & $\times$ & $\times$ & $1$ &\\[16pt]
$\shsc{10pt}{0.45}{\ydiagram{4,3,1}}$ & $\times$ & $1$ & $1$ & $\times$ & $2$ & $2$ &\\[16pt]
$\left[ 6, 4, 2\right]$ & $\times$ & $\times$ & $1$ & $\times$ & $1$ & $3$ &\\[16pt]
$\left[ 6, 5, 1 \right]$ & $\times$ & $\times$ & $1$ & $\times$ & $\times$ & $2$ &\\[16pt]
\end{tabular}
\end{center}
\end{table}

\clearpage
\newpage

\bibliography{references}
\bibliographystyle{apsrev4-1}

\end{document}

%% file: fig_aklt_state.tex
\begin{center}
\begin{tikzpicture}
\draw (-3.4,0.6) node[]{(a)};
\draw[white] (10.0,0.6) node[]{(a)};
\draw (0,0) ellipse(6mm and 3mm);
\draw (0.0,0.5) node[]{$\phys$};
\draw[fill] (-0.24,0.0) circle(1.2mm);
\draw[fill] (0.24,0.0) circle(1.2mm);
\draw[line width=1pt] (0.24,0.0) -- (1.76,0.0);
\draw (2.0,0.0) ellipse(6mm and 3mm);
\draw (2.0,0.5) node[]{$\phys$};
\draw[fill] (1.76,0.0) circle(1.2mm);
\draw[fill] (2.24,0.0) circle(1.2mm);
\draw[line width=1pt] (2.24,0.0) -- (3.76,0.0);
\draw (4.0,0.0) ellipse(6mm and 3mm);
\draw (4.0,0.5) node[]{$\phys$};
\draw[fill] (3.76,0.0) circle(1.2mm);
\draw[fill] (4.24,0.0) circle(1.2mm);
\draw[line width=1.4pt] (4.24,0.0) -- (5.76,0.0);
\draw (6.0,0.0) ellipse(6mm and 3mm);
\draw (6.0,0.5) node[]{$\phys$};
\draw[fill] (5.76,0.0) circle(1.2mm);
\draw[fill] (6.24,0.0) circle(1.2mm);
\draw[->] (-0.2,-0.8) to [out=170,in=-90] (-0.24,-0.15);
\draw[->] (0.2,-0.8) to [out=10,in=-90] (0.24,-0.15);
\draw (0.0,-0.8) node[]{$\vir$};
\draw[->] (1.8,-0.8) to [out=170,in=-90] (1.76,-0.15);
\draw[->] (2.2,-0.8) to [out=10,in=-90] (2.24,-0.15);
\draw (2.0,-0.8) node[]{$\vir$};
\draw[->] (3.8,-0.8) to [out=170,in=-90] (3.76,-0.15);
\draw[->] (4.2,-0.8) to [out=10,in=-90] (4.24,-0.15);
\draw (4.0,-0.8) node[]{$\vir$};
\draw[->] (5.8,-0.8) to [out=170,in=-90] (5.76,-0.15);
\draw[->] (6.2,-0.8) to [out=10,in=-90] (6.24,-0.15);
\draw (6.0,-0.8) node[]{$\vir$};
\end{tikzpicture}
\end{center}

%% file: fig_CGC.tex
\begin{center}
\begin{tikzpicture}
\draw (-2.4,1.5) node[]{(b)};
\draw[white] (3.5,1.5) node[]{(b)};
\draw[white] (3.5,-0.5) node[]{(b)};
\draw[fill] (0.866,0.5) -- (0.0,0.0) -- (0.0,1.0) -- (0.866,0.5);
\draw (-0.4,0.2) -- (0.0,0.2);
\draw (-0.4,0.8) -- (0.0,0.8);
\draw (0.866,0.5) -- (1.266,0.5);
\draw (-0.4,0.1) node[left]{$(\vir_1, \mathfrak{f}_a)$};
\draw (-0.4,0.9) node[left]{$(\vir_2, \mathfrak{g}_b)$};
\draw (1.266,0.5) node[right]{$(\phys, \mathfrak{h}_{\sigma})$};
\end{tikzpicture}
\end{center}

%% file: fig_aklt_state_mps.tex
\begin{center}
\begin{tikzpicture}
\draw (-1.2,1.0) node[]{(c)};
\draw[white] (5.2,1.0) node[]{(c)};
\draw[white] (5.2,-1.0) node[]{(c)};
\draw (-0.5,0.0) -- (4.7,0.0);
\draw (-0.5,0.0) to [out=180,in=-180] (-0.5,-0.5);
\draw (-0.5,-0.5) -- (4.7,-0.5);
\draw (4.7,0.0) to [out=0,in=0] (4.7,-0.5);
\draw[style={decorate, decoration=snake}] (0.0,0.0) -- (0.0,0.8);
\draw[style={decorate, decoration=snake}] (1.4,0.0) -- (1.4,0.8);
\draw[style={decorate, decoration=snake}] (2.8,0.0) -- (2.8,0.8);
\draw[style={decorate, decoration=snake}] (4.2,0.0) -- (4.2,0.8);
\draw[fill] (0.7,0.0) circle(1mm);
\draw[fill] (2.1,0.0) circle(1mm);
\draw[fill] (3.5,0.0) circle(1mm);
\draw[fill] (2.1,-0.5) circle(1mm);
\end{tikzpicture}
\end{center}

%% file: fig_aklt_ortho.tex
\begin{center}
\begin{tikzpicture}
\draw (-1.4,1.1) node[]{(d)};
\draw[white] (4.6,1.1) node[]{(d)};
\draw[white] (4.6,-1.3) node[]{(d)};
\draw (-0.2,0.0) -- (1.8,0.0);
\draw[style={decorate, decoration=snake}] (0.0,0.0) -- (0.0,0.6);
\draw[fill] (0.4,0.0) circle(0.6mm);
\draw[style={decorate, decoration=snake}] (0.8,0.0) -- (0.8,0.6);
\draw[fill] (1.2,0.0) circle(0.6mm);
\draw[style={decorate, decoration=snake}] (1.6,0.0) -- (1.6,0.6);
\draw (-0.2,0.0) to [out=180,in=-180] (-0.2,-0.3);
\draw (1.8,0.0) to [out=0,in=0] (1.8,-0.3);
\draw (-0.2,-0.3) -- (0.6,-0.3);
\draw (1.8,-0.3) -- (1.0,-0.3);
\draw (0.6,-0.3) to [out=0,in=90] (0.7,-0.5);
\draw (1.0,-0.3) to [out=180,in=90] (0.9,-0.5);
\draw[fill] (0.8,-0.8) -- (0.62,-0.5) -- (0.98,-0.5) -- (0.8,-0.8);
\draw[line width=1pt] (0.8,-0.7) -- (0.8,-1.0);
\end{tikzpicture}
\end{center}

%% file: fig_mps_H.tex
\begin{center}
\begin{tikzpicture}
\draw (-3.2,0.6) node[]{(e)};
\draw[white] (3.8,0.6) node[]{(e)};
\draw (-2.6,-0.85) node[anchor=west]{\large $h \ = \  \mathds{1} \ - $};
\draw (-0.2,0.0) -- (1.8,0.0);
\draw[style={decorate, decoration=snake}] (0.0,0.0) -- (0.0,0.4);
\draw[fill] (0.4,0.0) circle(0.6mm);
\draw[style={decorate, decoration=snake}] (0.8,0.0) -- (0.8,0.4);
\draw[fill] (1.2,0.0) circle(0.6mm);
\draw[style={decorate, decoration=snake}] (1.6,0.0) -- (1.6,0.4);
\draw (-0.2,0.0) to [out=180,in=-180] (-0.2,-0.3);
\draw (1.8,0.0) to [out=0,in=0] (1.8,-0.3);
\draw (-0.2,-0.3) -- (0.6,-0.3);
\draw (1.8,-0.3) -- (1.0,-0.3);
\draw (0.6,-0.3) to [out=0,in=90] (0.7,-0.5);
\draw (1.0,-0.3) to [out=180,in=90] (0.9,-0.5);
\draw[fill] (0.8,-0.8) --(0.62,-0.5) -- (0.98,-0.5) -- (0.8,-0.8);
\draw[line width=1pt] (0.8,-0.7) -- (0.8,-1.0);
\draw[fill] (0.8,-0.9) -- (0.62,-1.2) -- (0.98,-1.2) -- (0.8,-0.9);
\draw (0.7,-1.2) to [out=-90,in=0] (0.6,-1.4);
\draw (0.9,-1.2) to [out=-90,in=180] (1.0,-1.4);
\draw (-0.2,-1.4) -- (0.6,-1.4);
\draw (1.8,-1.4) -- (1.0,-1.4);
\draw (-0.2,-1.4) to [out=180,in=-180] (-0.2,-1.7);
\draw (1.8,-1.4) to [out=0,in=0] (1.8,-1.7);
\draw (-0.2,-1.7) -- (1.8,-1.7);
\draw[style={decorate, decoration=snake}] (0.0,-1.7) -- (0.0,-2.1);
\draw[fill] (0.4,-1.7) circle(0.6mm);
\draw[style={decorate, decoration=snake}] (0.8,-1.7) -- (0.8,-2.1);
\draw[fill] (1.2,-1.7) circle(0.6mm);
\draw[style={decorate, decoration=snake}] (1.6,-1.7) -- (1.6,-2.1);
\end{tikzpicture}
\end{center}

%% file: fig_su3_adjkronadj2.tex
\begin{center}
\begin{tikzpicture}
\draw (-0.5,0.8) node{(a)};
\draw[white] (9.6,0.8) node{(a)};
\draw (0,0.4) node[anchor=west]{$\shsc{2pt}{0.45}{\ydiagram{2,1}} \otimes \shsc{2pt}{0.45}{\ydiagram{2,1}} \ = \ \bullet \, \oplus 2 \shsc{2pt}{0.45}{\ydiagram{2,1}} \oplus \shsc{0pt}{0.45}{\ydiagram{3}} \oplus \shsc{2pt}{0.45}{\ydiagram{3,3}} \oplus \shsc{2pt}{0.45}{\ydiagram{4,2}}$};
\end{tikzpicture}
\end{center}

%% file: fig_su3_content_irreps2.tex
\begin{center}
\begin{tikzpicture}
\draw (-0.4,0.8) node[]{(b)};
\draw[white] (9.7,0.8) node[]{(b)};
\draw (0.0,0.4) node[anchor=west]{$\yset{U} = \left\{ \bullet, \, \shsc{2pt}{0.45}{\ydiagram{2,1}}, \, \shsc{0pt}{0.45}{\ydiagram{3}},  \, \shsc{2pt}{0.45}{\ydiagram{3,3}}, \, \shsc{2pt}{0.45}{\ydiagram{4,2}} \right\}$};
\end{tikzpicture}
\end{center}

%% file: fig_spin1_sets.tex
\begin{tikzpicture}[scale=1]
\draw (0.0,0.0) ellipse(2.6cm and 1.6cm);
\draw (-1.0,0.0) ellipse(1.2cm and 1.2cm);
\draw (2.7,0.9) node[]{$\phys^{\otimes 3}$};
\draw (-1.7,0.5) node[]{$\mathcal{K}$};
\draw (1.5,-0.8) node[]{$\mathcal{Q}$};
\draw (-1.0,0.6) ellipse(4mm and 4mm);
\draw (-1.0,0.6) node[]{\scalebox{0.6}{$K^{S=0}$}};
\draw (-1.5196,-0.3) ellipse(4mm and 4mm);
\draw (-1.5196,-0.3) node[]{\scalebox{0.6}{$K^{S=1}$}};
\draw (-0.48038,-0.3) ellipse(4mm and 4mm);
\draw (-0.48038,-0.3) node[]{\scalebox{0.6}{$K^{S=2}$}};
\draw (0.6,0.9) ellipse(4mm and 4mm);
\draw (0.6,0.9) node[]{\scalebox{0.6}{$Q^{S=1}_1$}};
\draw (0.6,-0.9) ellipse(4mm and 4mm);
\draw (0.6,-0.9) node[]{\scalebox{0.6}{$Q^{S=1}_2$}};
\draw (0.9,0.0) ellipse(4mm and 4mm);
\draw (0.9,0.0) node[]{\scalebox{0.6}{$Q^{S=2}$}};
\draw (1.9,0.0) ellipse(4mm and 4mm);
\draw (1.9,0.0) node[]{\scalebox{0.6}{$Q^{S=3}$}};
\end{tikzpicture}

%% file: fig_allstates_cgc.tex
\begin{center}
\begin{tikzpicture}
\draw[fill] (0.5,-0.866) -- (0.0,0.0) -- (1.0,0.0) -- (0.5,-0.866);
\draw (0.2,0.0) -- (0.2,0.4);
\draw (0.8,0.0) -- (0.8,0.4);
\draw[line width=2pt] (0.5,-0.7) -- (0.5,-0.8);
\draw[line width=2pt] (0.5,-0.8) to [out=-90,in=90] (0.8,-1.35);
\draw (1.4,0.4) -- (1.4,-1.3);
\draw[fill] (1.1,-2.166) -- (0.6,-1.3) -- (1.6,-1.3) -- (1.1,-2.166);
\draw[line width=2pt] (1.1,-2.1) -- (1.1,-2.3);
\draw[line width=2pt] (1.1,-2.3) to [out=-90,in=90] (1.2,-2.5);
\draw[fill] (1.675,0.0) circle(0.5pt);
\draw[fill] (1.95,0.0) circle(0.5pt);
\draw[fill] (2.225,0.0) circle(0.5pt);
\draw[fill] (1.325,-2.575) circle(0.5pt);
\draw[fill] (1.45,-2.65) circle(0.5pt);
\draw[fill] (1.575,-2.725) circle(0.5pt);
\draw (2.5,0.4) -- (2.5,-3.0);
\draw[fill] (2.2,-3.866) -- (2.7,-3.0) -- (1.7,-3.0) -- (2.2,-3.866);
\draw[line width=2pt] (1.9,-3.0) to [out=90,in=-90] (1.7,-2.8);
\draw[line width=2pt] (2.2,-3.8) -- (2.2,-4.1);
\draw (0.2,0.4) node[above]{\scalebox{0.5}{$(\phys, \sigma_1)$}};
\draw (0.8,0.4) node[above]{\scalebox{0.5}{$(\phys, \sigma_2)$}};
\draw (1.4,0.4) node[above]{\scalebox{0.5}{$(\phys, \sigma_3)$}};
\draw (2.5,0.4) node[above]{\scalebox{0.5}{$(\phys, \sigma_l)$}};
\draw (2.2,-4.1) node[below]{\scalebox{0.5}{$\mathlarger{\mathlarger{\oplus}} \, (\beta, \mathfrak{t}^{\beta}_{\xi}, j)$}};
\end{tikzpicture}
\end{center}

%% file: fig_su3_adjkronadj.tex
\begin{center}
\begin{tikzpicture}
\draw (-0.6,0.4) node{(a)};
\draw[white] (11.0,0.4) node{(a)};
\draw (0,0) node[anchor=west]{$\shsc{2pt}{0.4}{\ydiagram{2,1}} \otimes \shsc{2pt}{0.4}{\ydiagram{2,1}} \ = \ \bullet \, \oplus 2 \shsc{2pt}{0.4}{\ydiagram{2,1}} \oplus \shsc{0pt}{0.4}{\ydiagram{3}} \oplus \shsc{2pt}{0.4}{\ydiagram{3,3}} \oplus \shsc{2pt}{0.4}{\ydiagram{4,2}}$};
\draw (-0.2,-1.0) node[anchor=west]{$\shsc{0pt}{0.4}{\ydiagram{3}} \otimes \shsc{0pt}{0.4}{\ydiagram{3}} \ = \ \shsc{2pt}{0.4}{\ydiagram{3,3}} \oplus \shsc{2pt}{0.4}{\ydiagram{4,2}} \oplus \shsc{2pt}{0.4}{\ydiagram{5,1}} \oplus \shsc{0pt}{0.4}{\ydiagram{6}}$};
\end{tikzpicture}
\end{center}

%% file: fig_su3_content_irreps.tex
\begin{center}
\begin{tikzpicture}
\draw (-0.6,0.6) node[]{(b)};
\draw[white] (11.0,0.6) node[]{(b)};
\draw (-0.4,0.0) node[anchor=west]{$\yset{A} = \left\{ \shsc{3pt}{0.4}{\ydiagram{3,3}}, \, \shsc{3pt}{0.4}{\ydiagram{4,2}}, \, \shsc{3pt}{0.4}{\ydiagram{5,1}}, \, \shsc{0pt}{0.4}{\ydiagram{6}} \right\}$};
\draw (7.0,0.0) node[anchor=west]{$\yset{K} = \left\{ \shsc{3pt}{0.4}{\ydiagram{3,3}}, \,  \shsc{3pt}{0.4}{\ydiagram{4,2}} \right\}$};
\draw (-0.4,-1.2) node[anchor=west]{$\yset{U} = \left\{ \bullet, \, \shsc{3pt}{0.4}{\ydiagram{2,1}}, \, \shsc{0pt}{0.4}{\ydiagram{3}}, \, \shsc{3pt}{0.4}{\ydiagram{3,3}},\, \shsc{3pt}{0.4}{\ydiagram{4,2}} \right\}$};
\draw (6.0,-1.2) node[anchor=west]{$\yset{Q} = \left\{ \shsc{3pt}{0.4}{\ydiagram{5,1}}, \, \shsc{0pt}{0.4}{\ydiagram{6}} \right\}$};
\end{tikzpicture}
\end{center}

%% file: fig_su3_sets.tex
\begin{center}
\begin{tikzpicture}
\draw (-4.5,1.3) node[]{(c)};
\draw[white] (7.1,1.3) node[]{(c)};
\draw (0.0,0.0) ellipse(28mm and 16mm);
\draw (3.0,0.0) ellipse(28mm and 16mm);
\draw (5.8,-1.2) node[]{$\phys\otimes\phys$};
\draw (-2.0,-1.8) node[]{$\mathcal{U} = \vir\otimes\vir$};
\draw (-1.5,-0.6) circle(4mm);
\draw (-1.5,-0.6) node[]{\scalebox{0.6}{$U^{\bullet}$}};
\draw (-1.5,0.6) circle(4mm);
\draw (-1.5,0.6) node[]{\scalebox{0.6}{$U^{\shsc{2pt}{0.2}{\ydiagram{2,1}}}_1$}};
\draw (-0.4,0.6) circle(4mm);
\draw (-0.4,0.6) node[]{\scalebox{0.6}{$U^{\shsc{2pt}{0.2}{\ydiagram{2,1}}}_2$}};
\draw (-0.4,-0.6) circle(4mm);
\draw (-0.4,-0.6) node[]{\scalebox{0.6}{$U^{\shsc{2pt}{0.2}{\ydiagram{3}}}$}};
\draw (0.8,0.0) node[]{$\mathcal{K}$};
\draw (1.5,0.5) circle(4mm);
\draw (1.5,0.5) node[]{\scalebox{0.6}{$K^{\shsc{2pt}{0.2}{\ydiagram{3,3}}}$}};
\draw (1.5,-0.5) circle(4mm);
\draw (1.5,-0.5) node[]{\scalebox{0.6}{$K^{\shsc{2pt}{0.2}{\ydiagram{4,2}}}$}};
\draw (4.8,0.0) node[]{$\mathcal{Q}$};
\draw (3.8,0.7) circle(4mm);
\draw (3.8,0.7) node[]{\scalebox{0.6}{$Q^{\shsc{2pt}{0.2}{\ydiagram{5,1}}}$}};
\draw (3.8,-0.7) circle(4mm);
\draw (3.8,-0.7) node[]{\scalebox{0.6}{$Q^{\shsc{0pt}{0.2}{\ydiagram{6}}}$}};
\end{tikzpicture}
\end{center}

%% file: fig_su3_aklt_epsilontensor.tex
\begin{center}
\begin{tikzpicture}[scale=0.75]
\mysquare[](-1.5,2)(0.2);
\mysquare[](-1.2,2)(0.2);
\mysquare[](-0.9,2)(0.2);
\mysquare[](0,0)(0.2);
\mysquare[](0.3,0)(0.2);
\mysquare[](0.6,0)(0.2);
\mysquare[](1.5,2)(0.2);
\mysquare[](1.8,2)(0.2);
\mysquare[](2.1,2)(0.2);
\mysquare[](3,0)(0.2);
\mysquare[](3.3,0)(0.2);
\mysquare[](3.6,0)(0.2);
\mysquare[](4.5,2)(0.2);
\mysquare[](4.8,2)(0.2);
\mysquare[](5.1,2)(0.2);
\mysquare[](6,0)(0.2);
\mysquare[](6.3,0)(0.2);
\mysquare[](6.6,0)(0.2);
\mysquare[](7.5,2)(0.2);
\mysquare[](7.8,2)(0.2);
\mysquare[](8.1,2)(0.2);
\mysquare[](9,0)(0.2);
\mysquare[](9.3,0)(0.2);
\mysquare[](9.6,0)(0.2);
\mysquare[](10.5,2)(0.2);
\mysquare[](10.8,2)(0.2);
\mysquare[](11.1,2)(0.2);
\mysquare[](12,0)(0.2);
\mysquare[](12.3,0)(0.2);
\mysquare[](12.6,0)(0.2);
\draw[fill] (0.4,1.1) circle(1.5mm);
\draw[fill] (1.9,1.1) circle(1.5mm);
\draw[fill] (3.4,1.1) circle(1.5mm);
\draw[fill] (4.9,1.1) circle(1.5mm);
\draw[fill] (6.4,1.1) circle(1.5mm);
\draw[fill] (7.9,1.1) circle(1.5mm);
\draw[fill] (9.4,1.1) circle(1.5mm);
\draw[fill] (10.9,1.1) circle(1.5mm);
\draw (1.9,1.1) -- (1.9,2);
\draw (4.9,1.1) -- (4.9,2);
\draw (7.9,1.1) -- (7.9,2);
\draw (10.9,1.1) -- (10.9,2);
\draw (0.4,0.2) -- (0.4,1.1);
\draw (3.4,0.2) -- (3.4,1.1);
\draw (6.4,0.2) -- (6.4,1.1);
\draw (9.4,0.2) -- (9.4,1.1);
\draw (1.9,1.1) -- (0.7,0.2);
\draw (4.9,1.1) -- (3.7,0.2);
\draw (7.9,1.1) -- (6.7,0.2);
\draw (10.9,1.1) -- (9.7,0.2);
\draw (1.9,1.1) -- (3.1,0.2);
\draw (4.9,1.1) -- (6.1,0.2);
\draw (7.9,1.1) -- (9.1,0.2);
\draw (10.9,1.1) -- (12.1,0.2);
\draw (0.4,1.1) -- (1.6,2);
\draw (3.4,1.1) -- (4.6,2);
\draw (6.4,1.1) -- (7.6,2);
\draw (9.4,1.1) -- (10.6,2);
\draw (0.4,1.1) -- (-0.8,2);
\draw (3.4,1.1) -- (2.2,2);
\draw (6.4,1.1) -- (5.2,2);
\draw (9.4,1.1) -- (8.2,2);
\draw (0.4,0.1) ellipse (7mm and 3mm);
\draw (3.4,0.1) ellipse (7mm and 3mm);
\draw (6.4,0.1) ellipse (7mm and 3mm);
\draw (9.4,0.1) ellipse (7mm and 3mm);
\draw (12.4,0.1) ellipse (7mm and 3mm);
\draw (-1.1,2.1) ellipse (7mm and 3mm);
\draw (1.9,2.1) ellipse (7mm and 3mm);
\draw (4.9,2.1) ellipse (7mm and 3mm);
\draw (7.9,2.1) ellipse (7mm and 3mm);
\draw (10.9,2.1) ellipse (7mm and 3mm);
\draw[decorate,decoration={snake}] (-1.1,2.4) -- (-1.1,3.0);
\draw[decorate,decoration={snake}] (1.9,2.4) -- (1.9,3.0);
\draw[decorate,decoration={snake}] (4.9,2.4) -- (4.9,3.0);
\draw[decorate,decoration={snake}] (7.9,2.4) -- (7.9,3.0);
\draw[decorate,decoration={snake}] (10.9,2.4) -- (10.9,3.0);
\draw[decorate,decoration={snake}] (0.4,-0.2) -- (0.4,-0.8);
\draw[decorate,decoration={snake}] (3.4,-0.2) -- (3.4,-0.8);
\draw[decorate,decoration={snake}] (6.4,-0.2) -- (6.4,-0.8);
\draw[decorate,decoration={snake}] (9.4,-0.2) -- (9.4,-0.8);
\draw[decorate,decoration={snake}] (12.4,-0.2) -- (12.4,-0.8);
\end{tikzpicture}
\end{center}

%% file: fig_su3_aklt_irrep420_33bar.tex
\begin{center}
\scalebox{0.75}{
\begin{tikzpicture}
\draw[fill,white] (11.5,1.05) circle(0.5mm);
\myrectangle[](0,0.0)(0.2);
\myrectangle[](0,0.6)(0.2);
\mysquare[](0,1.3)(0.2);
\mysquare[](0,1.9)(0.2);
\mysquare[](1.5,0.1)(0.2);
\mysquare[](1.5,0.7)(0.2);
\myrectangle[](1.5,1.2)(0.2);
\myrectangle[](1.5,1.8)(0.2);
\myrectangle[](3,0.0)(0.2);
\myrectangle[](3,0.6)(0.2);
\mysquare[](3,1.3)(0.2);
\mysquare[](3,1.9)(0.2);
\mysquare[](4.5,0.1)(0.2);
\mysquare[](4.5,0.7)(0.2);
\myrectangle[](4.5,1.2)(0.2);
\myrectangle[](4.5,1.8)(0.2);
\myrectangle[](6,0.0)(0.2);
\myrectangle[](6,0.6)(0.2);
\mysquare[](6,1.3)(0.2);
\mysquare[](6,1.9)(0.2);
\mysquare[](7.5,0.1)(0.2);
\mysquare[](7.5,0.7)(0.2);
\myrectangle[](7.5,1.2)(0.2);
\myrectangle[](7.5,1.8)(0.2);
\myrectangle[](9,0.0)(0.2);
\myrectangle[](9,0.6)(0.2);
\mysquare[](9,1.3)(0.2);
\mysquare[](9,1.9)(0.2);
\mysquare[](10.5,0.1)(0.2);
\mysquare[](10.5,0.7)(0.2);
\myrectangle[](10.5,1.2)(0.2);
\myrectangle[](10.5,1.8)(0.2);
\draw[line width=1pt] (0.25,0.2) -- (1.45,0.2);
\draw[line width=1pt] (0.25,1.4) -- (1.45,1.4);
\draw[line width=1pt] (1.75,0.8) -- (2.95,0.8);
\draw[line width=1pt] (1.75,2.0) -- (2.95,2.0);
\draw[line width=1pt] (3.25,0.2) -- (4.45,0.2);
\draw[line width=1pt] (3.25,1.4) -- (4.45,1.4);
\draw[line width=1pt] (4.75,0.8) -- (5.95,0.8);
\draw[line width=1pt] (4.75,2.0) -- (5.95,2.0);
\draw[line width=1pt] (6.25,0.2) -- (7.45,0.2);
\draw[line width=1pt] (6.25,1.4) -- (7.45,1.4);
\draw[line width=1pt] (7.75,0.8) -- (8.95,0.8);
\draw[line width=1pt] (7.75,2.0) -- (8.95,2.0);
\draw[line width=1pt] (9.25,0.2) -- (10.45,0.2);
\draw[line width=1pt] (9.25,1.4) -- (10.45,1.4);
\draw (0.1,1.05) ellipse(4mm and 14mm);
\draw (1.6,1.05) ellipse(4mm and 14mm);
\draw (3.1,1.05) ellipse(4mm and 14mm);
\draw (4.6,1.05) ellipse(4mm and 14mm);
\draw (6.1,1.05) ellipse(4mm and 14mm);
\draw (7.6,1.05) ellipse(4mm and 14mm);
\draw (9.1,1.05) ellipse(4mm and 14mm);
\draw (10.6,1.05) ellipse(4mm and 14mm);
\end{tikzpicture}}
\end{center}

%% file: fig_su4_irreps.tex
\begin{tikzpicture}
\draw (-1.8,1.4) node[anchor=west]{(a)};
\draw[white] (4.2,1.4) node[anchor=west]{(a)};
\draw (-1.0,0.4) node[anchor=west]{\small $\phys = $};
\myrectangle[](0.0,0.0)(0.2);
\myrectangle[](0.0,0.4)(0.2);
\myrectangle[](0.2,0.0)(0.2);
\myrectangle[](0.2,0.4)(0.2);
\draw [decorate, decoration={brace,amplitude=3pt, mirror, raise=3ex}]
  (0.0,0.0) -- (0.0,0.8) node[midway, xshift=3em]{\tiny $N/2$};
\draw (1.4,0.4) node[anchor=west]{\small $, \quad \vir = $};
\myrectangle[](2.9,0.0)(0.2);
\myrectangle[](2.9,0.4)(0.2);
\draw [decorate, decoration={brace,amplitude=3pt, mirror, raise=4ex}]
  (2.5,0.0) -- (2.5,0.8) node[midway, xshift=3em]{\tiny $N/2$};
\draw (5.0,1.4) node[anchor=west]{(b)};
\draw[white] (11.0,1.4) node[anchor=west]{(b)};
\draw (5.8,0.4) node[anchor=west]{\small $\phys = $};
\mysquare[](7.0,0.8)(0.2);
\myrectangle[](6.8,0.6)(0.2);
\myrectangle[](6.8,0.2)(0.2);
\myrectangle[](6.8,-0.2)(0.2);
\mysquare[](6.8,-0.4)(0.2);
\draw [decorate, decoration={brace,amplitude=4pt, mirror, raise=3ex}]
  (6.8,-0.4) -- (6.8,1.0) node[midway, xshift=3em]{\tiny $N-1$};
\draw (8.3,0.4) node[anchor=west]{\small $, \quad  \vir = $};
\myrectangle[](9.8,0.0)(0.2);
\myrectangle[](9.8,0.4)(0.2);
\draw [decorate, decoration={brace,amplitude=3pt, mirror, raise=2ex}]
  (9.8,0.0) -- (9.8,0.8) node[midway, xshift=2em]{\tiny $N/2$};
\end{tikzpicture}